\documentclass[12pt]{article}
\usepackage{changes}
\usepackage[T1]{fontenc}
\usepackage{verbatim}
\usepackage{float}
\usepackage{amsthm}
\usepackage{amsmath}
\usepackage{amssymb}
\usepackage{graphicx}
\usepackage{color}
\usepackage{url}
\usepackage{caption}
\usepackage{subcaption}
\usepackage{mathtools} 
\usepackage{stackrel} 

\newcommand{\E}{\mathbb{E}}

\newcommand{\II}{\mathcal{I}}

\newcommand{\mb}{\mathbf}
\newcommand{\mk}{\mathfrak}
\newcommand{\mc}{\mathcal}
\newcommand*\Bell{\ensuremath{\boldsymbol\ell}}
\makeatletter

\newcommand{\reals}{\mathbb{R}}

\newcommand{\tamir}{x}

\newcommand{\RNN}{\mathbb{R}^{N\times N}}

\newcommand{\be}
{\begin{equation}}
\newcommand{\ee}
{\end{equation}}

\renewcommand{\P}{\mathrm{Prob}}

\theoremstyle{plain}
\newtheorem{thm}{\protect\theoremname}[section]
\theoremstyle{definition}

\theoremstyle{remark}

\theoremstyle{plain}

\newtheorem*{lem*}{Lemma}
\theoremstyle{remark}

\theoremstyle{plain}

\theoremstyle{plain}
\newtheorem{proposition}[thm]{\protect\propositionname}
\providecommand{\claimname}{Claim}
\providecommand{\definitionname}{Definition}
\providecommand{\lemmaname}{Lemma}
\providecommand{\remarkname}{Remark}
\providecommand{\theoremname}{Theorem}
\providecommand{\corollaryname}{Corollary}
\providecommand{\propositionname}{Proposition}
\numberwithin{equation}{section}

\usepackage{authblk}

\usepackage[margin=.6in]{geometry}

\newenvironment{myquote}[1]%
{\list{}{\leftmargin=#1\rightmargin=#1}\item[]}%
{\endlist}

\allowdisplaybreaks

\begin{document}


\title{Toward single particle reconstruction without particle picking: Breaking the detection limit}

\author[a]{Tamir Bendory}
\author[b]{Nicolas Boumal} 
\author[c]{William Leeb}
\author[d]{Eitan Levin}
\author[e]{Amit Singer}

\affil[a]{The School of Electrical Engineering, Tel Aviv University, Tel Aviv, Israel}
\affil[b]{Institute of Mathematics, Ecole Polytechnique F\'ed\'erale
 de Lausanne (EPFL), Lausanne, Switzerland}
\affil[c]{School of Mathematics, University of
	Minnesota, Minneapolis, Minnesota, USA }
\affil[d]{Department of Computing and Mathematical Sciences, California Institute of Technology, Pasadena, California, USA }
\affil[e]{The Program in Applied and Computational Mathematics and  Department of Mathematics, Princeton University, Princeton, NJ, USA}

\maketitle

\begin{abstract}
Single-particle cryo-electron microscopy (cryo-EM) has recently joined X-ray crystallography
and NMR spectroscopy as a high-resolution structural method to resolve
 biological macromolecules.
In a cryo-EM experiment, the microscope produces images called micrographs. Projections of the molecule of interest are embedded in the micrographs at unknown locations, and under unknown viewing directions. Standard imaging techniques first locate these projections (detection) and then reconstruct the 3-D structure from them. Unfortunately, high noise levels hinder detection. When reliable detection is rendered impossible, the standard techniques fail. This is a problem, especially for small molecules. 
In this paper, we pursue a radically different approach: we contend that the structure could, in principle, be reconstructed directly from the micrographs, without intermediate detection. 
The aim is to bring  small molecules within reach for cryo-EM.
To this end, we design an autocorrelation analysis technique that allows to go directly from the micrographs to the sought structures. This involves only one pass over the micrographs, allowing online, streaming processing for large experiments. 
We show numerical results and  discuss challenges that lay ahead to turn this proof-of-concept into a complementary approach to state-of-the-art algorithms. 
\end{abstract}

\section{Introduction}

\begin{figure}[t]
	\centering
	\begin{subfigure}[h]{0.32\linewidth}
		\centering
		\includegraphics[width=1\linewidth]{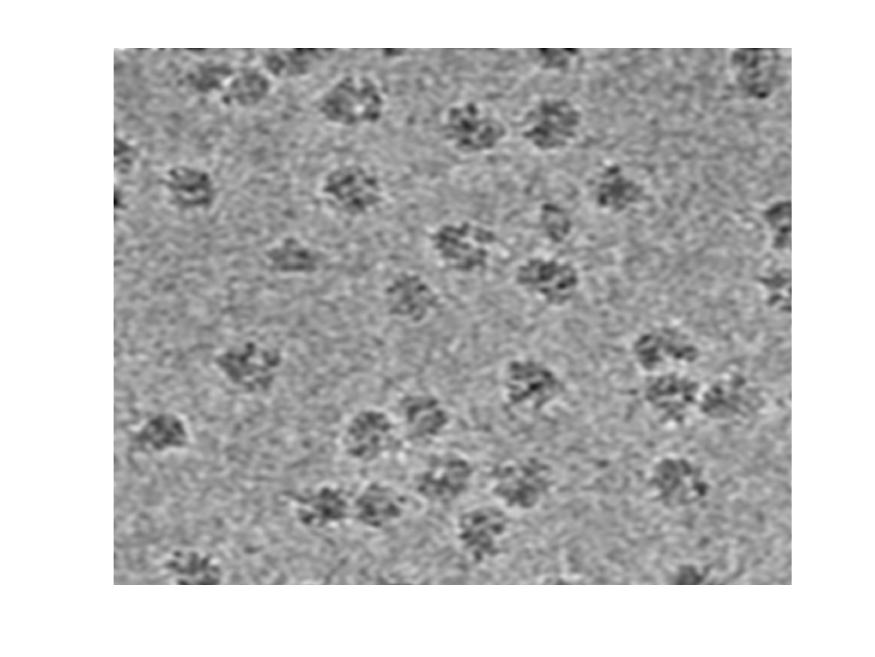}
		\caption{\label{fig:10028} \texttt{EMPIAR 10028}}
	\end{subfigure}%
	\begin{subfigure}[h]{0.32\linewidth}
		\centering
		\includegraphics[width=1\linewidth]{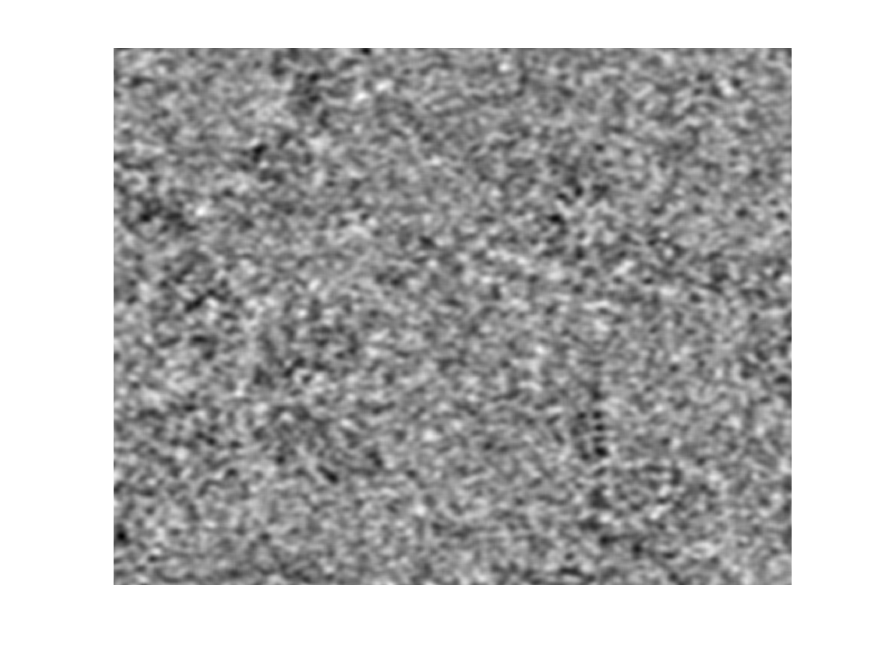}
		\caption{\label{fig:10061} \texttt{EMPIAR 10061}}
	\end{subfigure}
	\begin{subfigure}[h]{0.32\linewidth}
	\centering
	\includegraphics[width=1\linewidth]{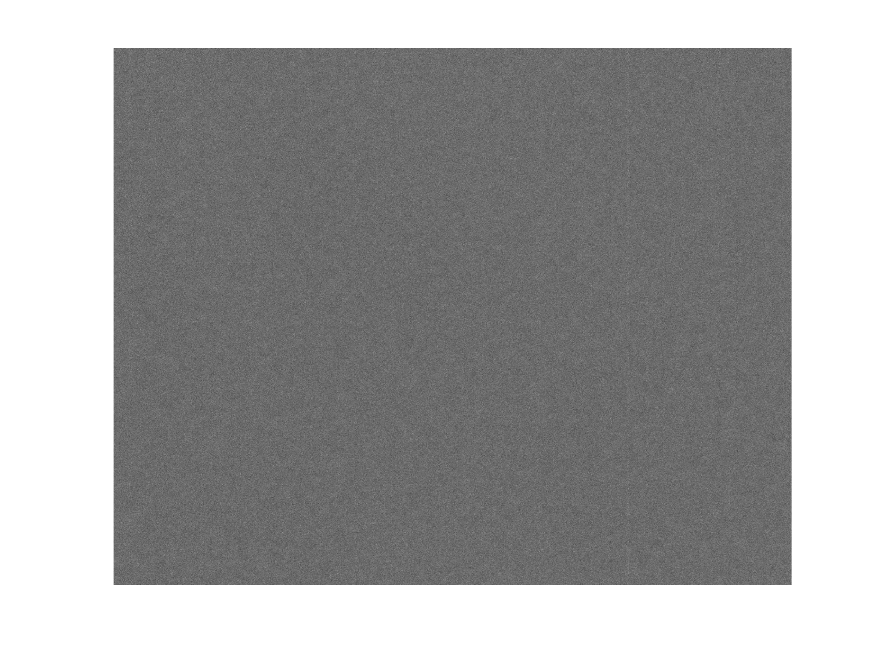}
	\caption{\label{fig:10249} \texttt{EMPIAR 10249}}
\end{subfigure}
	\caption{\label{fig:exp_micro_example} Excerpts from three data sets from the EMPIAR repository~\cite{iudin2016empiar}: \texttt{EMPIAR 10028} (Plasmodium Falciparum 80S ribosome)~\cite{wong2014cryo},   \texttt{EMPIAR 10061} ($\beta$-Galactosidase)~\cite{bartesaghi20152}, and \texttt{EMPIAR 10249} (Alcohol Dehydrogenase)~\cite{herzik2019high}. 	
	Detecting particles in the \texttt{EMPIAR 10028} data set (left panel) is rather easy as the particles in the micrograph, corresponding to a large ribosome with a molecular weight of 4 MDa, are clearly visible. 
	Doing so in the \texttt{EMPIAR 10061} data set (middle panel) is harder, as the $\beta$-Galactosidase is a smaller protein with a molecular weight of 465 kDa, but some particles can still be identified visually.
	The particles in the \texttt{EMPIAR 10249} data set (right panel) are tomographic projections of a small molecule that weighs  82 kDa, yielding  low SNR.
	We note that the differences in particle contrast are also attributed to the fact that the \texttt{EMPIAR 10249} data set  was obtained with a microscope operating at 200 keV, whereas the \texttt{EMPIAR 10028} and  \texttt{EMPIAR 10061} data sets were taken at 300 keV.      
	 In this work, we propose a method that aims to recover molecular structures when the SNR is very low, even below the SNR of the \texttt{EMPIAR 10249} data set.	}
\end{figure}

Cryo-electron microscopy (cryo-EM) is an imaging technique in structural biology used for single particle reconstruction  of macromolecules.
In a cryo-EM experiment, biological samples are rapidly frozen in a thin layer of vitreous ice. 
The microscope produces a 2-D tomographic image of the samples embedded in the ice, called a \emph{micrograph}. Each micrograph contains tomographic projections of the samples at unknown locations and under unknown viewing directions. Figure~\ref{fig:exp_micro_example} presents  three experimental micrographs.
 The goal is to construct 3-D models of the molecular structure from the micrographs~\cite{frank2006three,bendory2020single,singer2020computational}.

The signal-to-noise ratio (SNR) of the tomographic projections in the micrographs is a function of two dominating factors. On the one hand, the SNR is a function of the electron dose. To keep radiation damage within acceptable bounds, the dose must be kept low, which leads to high noise levels. On the other hand, the SNR is a function of the molecule size. The smaller the molecules, the fewer detected electrons carry information about them.

All methods currently in use split the reconstruction procedure into two main  stages.
The first stage consists in extracting the various particle projections from the micrographs. This stage is called \emph{particle picking}~\cite{scheres2015semi,heimowitz2018apple,bepler2019positive,eldar2020klt}. The second stage aims to construct a 3-D model of the molecular structure from these projections. The quality of the reconstruction eventually hinges on the quality of the particle picking stage.
Figure~\ref{fig:10028} shows an example of a data set of a large ribosome with a  molecular weight of 4MDa, where the particles in the micrograph are clearly visible, and thus particle picking is rather easy.
The particles in Figure~\ref{fig:10061} correspond to a smaller protein with a molecular weight of 465 kDa, but some particles can still be identified visually.
The particles in Figure~\ref{fig:10249} are tomographic projections of a small molecule that weighs  82 kDa, yielding  low SNR. 

%

Crucially, it can be shown that reliable detection of individual particles is impossible below a certain critical SNR. This fact has been recognized early on by the cryo-EM community. In particular, in an influential paper from 1995, Nobel laureate Richard Henderson~\cite{henderson1995potential} investigates the following questions:
\begin{myquote}{16pt}
	\emph{For the purposes of this review, I would like to ask the question: what is the smallest size of free-standing molecule whose structure can in principle be determined by phase-contrast electron microscopy? Given what has already been demonstrated in published work, this reduces to the question: what is the smallest size of molecule for which it is possible to determine from images of unstained molecules the five parameters needed to define accurately its orientation (three parameters) and position (two parameters) so that averaging can be performed?}
\end{myquote}
In that paper and in others that followed (e.g.,~\cite{glaeser1999electron}), it was established that particle picking is impossible for molecules below a certain weight (below $\sim$40kDa). 
Joachim Frank voices a similar observation in his 2017 Nobel Prize lecture: ``\emph{Using the ribosome as an example, it became clear from the formula we obtained that the single-particle approach to structure research was indeed feasible for molecules of sufficient size: Particle Size > 3/[Contrast$^2$ $\times$ Resolution (as length) $\times$ Critical Electron Dose]}''~\cite{frank2018single}. 
As these two leaders of the cryo-EM field point out, it is impossible to reconstruct small molecules by any of the existing computational pipelines for single particle analysis in cryo-EM, because the particles themselves cannot be picked from the micrographs. 
Given the fact that the vast majority of the proteins
that make up the mammalian proteome weigh less than 100 kDa, whereas the molecular weights of
almost all biological macromolecules whose structures have been determined using cryo-EM is greater
than 100 kDa, new methodologies targeting macromolecules in this size are of pivotal importance~\cite{lander2021conquer}.
In addition, recovering small structures is crucial for  structure-guided drug design~\cite{scapin2018cryo}.

Recovering smaller molecular structures using cryo-EM is an active research effort~\cite{zhang2019cryo,wu2020low,bai2021seeing}, mostly
focused on sample preparation techniques and hardware developments, such as Volta phase plates~\cite{khoshouei2017cryo,liang2017phase}, laser phase plates~\cite{schwartz2019laser}, and scaffolding cages~\cite{liu2018near}, as well as steady improvements in the data processing pipeline.
Despite this progress, detecting small molecules in the micrographs remains a challenge (see Figure~\ref{fig:10249}).
We note that nuclear magnetic resonance (NMR) spectroscopy and X-ray crystallography are well suited to reconstruct small molecules. Yet, cryo-EM has a lot to offer even for molecules with already known structures obtained via NMR spectroscopy or X-ray crystallography, because these methods have limited ability to distinguish conformational variability~\cite{sorzano2019survey}.

In this paper, we argue that there is a gap between the two questions in Henderson's quoted excerpt above, and that one may be able to exploit it to design better reconstruction algorithms.
Specifically, the impossibility of particle picking does not necessarily imply impossibility of particle reconstruction.
Indeed, the aim is only to reconstruct the molecule: estimating the locations of the particles in the micrograph is merely a helpful intermediate stage when it can be done. 
Our main message is that, while a molecule's size may limit our ability to pick particles, that does not necessarily translate into a limit on our ability to reconstruct the molecule.

In order to recover the 3-D structure, we use autocorrelation analysis that relates the autocorrelations of the micrographs to the parameters describing the 3-D model. 
For any noise level, the autocorrelations of the micrographs can be estimated to any desired accuracy, provided we  observe sufficiently many micrographs. 
The autocorrelations of the micrographs are straightforward to compute and require only one pass over the data.
To estimate the 3-D structure itself from the estimated autocorrelations, we  formulate and solve a nonlinear inverse problem. 
In particular, we show that the first order autocorrelation is just the average pixel value, the second-order autocorrelation is effectively a 1-D radial function, and the third-order is a  3-D function. We may therefore hope (but not guarantee) to recover the 3-D volume from the third-order autocorrelation, as it provides sufficiently many equations to solve for the parameters.
Importantly, there is no need to detect individual projection images. 
We show a few numerical examples and outline the future developments required to make this method applicable to experimental  data.

Another interesting feature of the described approach pertains to model bias, whose importance in cryo-EM was stressed by a number of authors~\cite{shatsky2009method,van1992correlation,henderson2013avoiding,van2013finding}. In the classical ``Einstein from noise'' experiment, multiple realizations of pure noise are aligned to a picture of Einstein using template matching and then averaged. In~\cite{shatsky2009method}, it was shown that the averaged noise rapidly becomes remarkably similar to the Einstein template. In the context of cryo-EM, this experiment exemplifies how prior assumptions about the particles may influence the reconstructed structure. This model bias is common to all particle picking methods based on template matching. 
In our approach, we do not attempt to match a template to noisy data.
	Instead, we use an optimization algorithm to try to solve the inverse problem relating autocorrelations to the sought volume.
	This nonconvex problem may have spurious local minima, hence the algorithm's initialization (a form of template) may affect the outcome.
	However, we find empirically that this poses a lesser risk because we are able to obtain meaningful reconstructions from random initializations.

The rest of the paper is organized as follows. 
Section~\ref{sec:detection} shows that it is impossible to  detect the particle images in the micrograph  in extremely low SNR regimes.
In Section~\ref{sec:ac_cryo}, which is the main contribution of this paper, we develop an autocorrelation analysis framework to recover molecular structures directly from the micrograph, circumventing particle picking.  Section~\ref{sec:numerical_experiments} shows a few numerical experiments, and Section~\ref{sec:discussion} concludes the paper and delineates challenges that lay ahead to turn this proof-of-concept into a complementary approach to state-of-the-art algorithms.

\section{The detection limit} \label{sec:detection}

In the low SNR regime---even if the 3-D structure is known---one cannot reliably detect the projection images in the micrograph.
To support this claim, we consider a strictly simpler  problem: suppose an oracle identifies for us one patch in the micrograph that either contains a full signal occurrence (plus noise), or contains just noise. Our task is to determine which one it is.
The oracle further provides the true signal $x$, the noise variance $\sigma^2$, and the probability $q$ that we observe signal-plus-noise as opposed to pure noise.

This decision problem can be abstracted as follows:  We have two known vectors $\theta_0 = x$ and $\theta_1 = 0$ in $\mathbb{R}^L$. There is a random variable $\eta$ taking values 0 or 1 with probabilities $q$ and $1-q$, respectively. We observe a random vector $X \in \mathbb{R}^L$ (akin to an extract of the micrograph) with the following distribution: if $\eta = 0$, then $X \sim \mathcal N(\theta_0,\sigma^2I_L)$ and if $\eta = 1$, then $X \sim \mathcal N(\theta_1,\sigma^2I_L)$. 

We observe a realization of $X$, and our task is to estimate $\eta$. How reliably can this be done? If $q \geq 1/2$, the constant estimator $\hat{\eta} = 0$ is correct with probability $q$; likewise, if $q \leq 1/2$, the constant estimator $\hat{\eta} = 1$ is correct with probability $1-q$. The question is, can we do better than this?  We prove that, as $\sigma \to \infty$, the answer is no. The result is proved in Appendix~\ref{sec:proof_two_gauss}.

\begin{proposition} \label{prop:two_gauss}
	For any deterministic estimator $\hat{\eta }$ of $\eta$,
	\begin{equation}
		\lim_{\sigma \to \infty} \P[ \hat{\eta} = \eta  ] \leq \max(q, 1-q);
		\end{equation}
	that is, as the SNR deteriorates, the probability of success is no better than random chance.
\end{proposition}

Proposition~\ref{prop:two_gauss} implies that, at low SNR, particle picking is impossible: in order to estimate the 3-D structure, we must consider methods that aim to estimate the structure directly, without estimating the nuisance locations of the projection images  as an intermediate step.
In the next section, we develop an autocorrelation analysis technique for that purpose.

\section{Autocorrelation analysis for cryo-EM} \label{sec:ac_cryo}

\subsection{Autocorrelation analysis}

For a random signal $z\in\mathbb{R}^{N\times N}$, the autocorrelation of order $p$ is given by
\begin{equation}
	a_z^p[\Bell_1,\ldots,\Bell_{p-1}] = \E_z\left\{ \frac{1}{N^2}\sum_{\mb i}z[\mb i]z[\mb i+\Bell_1]\cdots z[\mb i+\Bell_{p-1}]\right\},
\end{equation}
where the expectation is taken with respect to the distribution of $z$,  the summation is for $\mb i$  ranging over the $N^2$ pixels of the signal, and $\Bell_1,\ldots,\Bell_{q-1}$ are two-dimensional integer shifts.
Indexing out of bounds is
zero-padded, that is, $z[\mb i] = 0$ for $\mb i$ out of the range $[0,N-1]\times [0,N-1]$.
In our case, $z$ will represent the micrographs, whose distribution is parameterized by the sought 3-D structure. 
In the next sections, we show that as $N\to\infty$, the first three autocorrelations of the micrograph (i.e., $p=1,2,3$) converge to explicit polynomial functions of the parameters describing the sought volume.
Then, we wish to estimate the volume  by finding a set of parameters which are consistent with the autocorrelations; this entails solving a non-convex optimization problem.

We mention that autocorrelation analysis for cryo-EM images was first proposed in 1980 by Zvi Kam~\cite{kam1980reconstruction}. However, Kam's method, as well as some of its recent extensions~\cite{levin20183d,sharon2020method}, assumes picked particles. 
To break this fundamental barrier, we suggest computing the autocorrelations of the  micrographs directly, completely bypassing particle picking. 

\subsection{Model and autocorrelation functions}

Let $\phi \colon \reals^3 \to \reals$ be the Coulomb potential representing the molecule we aim to recover. 
We assume that the molecule is smooth. 
Specifically, in spherical coordinates, its
3-D Fourier transform $\widehat\phi$ admits a finite expansion of the form
\begin{equation} \label{eq:volume_expansion_app} 
	\widehat \phi(ck, \theta, \varphi) = \sum_{\ell = 0}^{L_{\text{max}}}\sum_{m=-\ell}^{\ell}\sum_{s=1}^{S(\ell)}\tamir_{\ell, m, s}Y_{\ell}^m(\theta,\varphi)j_{\ell,s}(k), \quad k\leq 1,
\end{equation}
where $0<c\leq 0.5$ is the bandlimit in Fourier space (a standard assumption in cryo-EM) $\{S(\ell)\}$ are determined using the Nyquist criterion as described in~\cite{bhamre2017anisotropic,klug1972three}, $j_{\ell,s}$ is the normalized spherical Bessel functions given by
\[ j_{\ell, s}(k) = \frac{4}{|j_{\ell+1}(u_{\ell, s})|}j_{\ell}(u_{\ell,s} k),\]
$j_{\ell}$ is the spherical Bessel function of order $\ell$ and  $u_{\ell,s}$ is the $s$th positive zero of $j_{\ell}$. We use 
the complex spherical harmonics $Y_{\ell}^m$ defined by  
\[ Y_{\ell}^m(\theta,\varphi) := \sqrt{\frac{2\ell+1}{4\pi}\cdot\frac{(\ell-m)!}{(\ell+m)!}}P_{\ell}^m(\cos\theta)e^{\iota m\varphi},\]
where $P_{\ell}^m$ are the associated Legendre polynomials with the Condon-Shortley phase. 
Representing the molecule using a spherical harmonics expansion, akin to~\eqref{eq:volume_expansion_app}, is a ubiquitous practice in the cryo-EM literature, see for example~\cite{bandeira2020non}.
Sampling at the Nyquist rate dictates $c=1/2$~\cite{levin20183d}. 
Because $\phi$ is real-valued,  $\widehat \phi$ is conjugate-symmetric and thus the expansion coefficients satisfy $\tamir_{\ell, -m, s} = (-1)^{\ell+m}\overline{\tamir_{\ell,m,s}}$. Therefore, we only need to recover coefficients $\tamir_{\ell, m, s}$ with $m\geq 0$ in order to recover the molecular structure.

Let $I_{\omega}$ denote the tomographic projection obtained from viewing direction $\omega\in SO(3)$. 
The Fourier slice theorem states that the 2-D Fourier transform of a tomographic projection is equal to a 2-D slice of the  volume's 3-D Fourier transform~\cite{natterer2001mathematics}. Specifically, 

\begin{equation}\label{eq:projection_model}
	\widehat I_{\omega}(ck,\varphi) = \sum_{\ell,m,m',s}\tamir_{\ell,m,s}D_{m',m}^{\ell}(\omega)Y_{\ell}^{m'}\left(\frac{\pi}{2},\varphi\right)j_{\ell,s}(k),
\end{equation}
where $D_{m',m}^{\ell}(\omega)$ is a Wigner-D matrix. This implies that the projections are also $c$-bandlimited.
In practice, the projections are further affected by additional factors, such as the microscope's point spread function, which are neglected in this paper.

Let $\II\in\RNN$ denote a micrograph. We assume it consists of shifted copies of projections contaminated by additive white Gaussian noise:
\begin{equation}\label{eq:micrograph_model}
	\II = \sum_{t=1}^{M} I_{\omega_t}\ast\delta_{\mb s_t}+ \varepsilon, \quad \varepsilon~\sim\mathcal{N}(0,\sigma^2 I),
\end{equation}
where $\ast$ denotes convolution, and the viewing directions $\omega_t$ are assumed to be drawn from the uniform distribution over $\text{SO}(3)$.
We assume the projections are discretized on a Cartesian grid  of size $P\times P$ 
 and $\mb s_t\in [P,N-P]^2$ denotes the location of the upper-left corner of the $t$-th projection in the micrograph. 
 We impose a \emph{separation condition} so that any two projections are separated by at least $2P-1$ pixels between their upper-left corners,  in each direction.
Thus, their end points (in each direction) are necessarily separated by at least $P-1$ signal-free entries in the micrograph.

We stress that the assumptions of the generative model~\eqref{eq:micrograph_model} are unrealistic for experimental cryo-EM data. 	For example, in practice, the particles are not likely to satisfy the separation condition (but they do not overlap and thus are separated by at least $P-1$ pixels in each direction), and the distribution over $SO(3)$ is typically non-uniform; these assumptions were made to ease the analysis. 
Section~\ref{sec:discussion} discusses how our autocorrelation analysis technique can be extended to include these important  aspects  that are essential to  reach high-resolution reconstructions.

Define the $p$th (empirical) autocorrelation of $\II$ as
\begin{equation} \label{eq:Kth_autocorrelation}
	a^p_\II[\Bell_1,\ldots, \Bell_{p-1}] := \frac{1}{N^2}\sum_{\mb i}{\II[\mb i]}\II[\mb i+\Bell_1]\cdots \II[\mb i+\Bell_{p-1}],
\end{equation}
where the summation is for $\mb i $ ranging over the $N^2$  pixels of the micrograph,  using zero padding when indices exceed the micrograph's edge.
Computing the autocorrelations of the micrograph is straightforward. 
Let $\II_1,\ldots,\II_K$ denote a set of $K$ micrographs. 
Under the specified conditions, we show in the next section that the first three autocorrelations  of the micrographs are related to those of the projections by 
\begin{align} \label{eq:ac_micrographs}
	\lim_{K\to\infty}\frac{1}{K}\sum_{i=1}^Ka^p_{\II_i}[\Bell_1,\ldots,\Bell_{p-1}]  &= \gamma\left\langle a^p_{I_\omega}[\Bell_1,\ldots,\Bell_{p-1}]\right\rangle_{\omega} + b_p[\Bell_1,\ldots,\Bell_{p-1}], \\ p=1,2,3, \quad & \Bell_1,\ldots,\Bell_{p-1}\in[-(P-1),P-1]^2, \nonumber
\end{align}
where $\langle\cdot\rangle_\omega$ denotes averaging over all possible viewing directions $\omega$ with respect to the uniform measure, $\gamma = \frac{MP^2}{N^2}\in(0,1)$ is a scalar that encodes the density of the particle projections in the data, and $b_p$ is a bias term. 
Specifically,  $b_1 = 0$ and therefore the mean is unbiased. The bias term of the second-order autocorrelation  $b_2$ depends only on $\sigma^2$, the variance of the noise. Hence, if the noise level can be accurately estimated from the micrographs, this bias can be removed. 
Finally, the bias term of the third-order autocorrelation $b_3$ depends on the mean of the micrograph and $\sigma^2$.  Therefore, given sufficiently many projections, we can accurately estimate the quantities $\gamma\langle a^p_{I_{\omega}}\rangle_{\omega}$ directly from the micrographs. These quantities are functions of the unknown coefficients $\tamir_{l,m,s}$ and we could proceed to invert their relation. 

In practice, we want to leverage one more feature of the 3-D reconstruction problem.
Since all in-plane rotations of the micrographs are equally likely observations, 
it is desirable in~\eqref{eq:ac_micrographs} to average over all in-plane rotations as well.
This can be done efficiently using Prolate Spheroidal Wave Functions (PSWFs), which have been used before for cryo-EM data processing~\cite{landa2017steerable,landa2018steerable,sharon2020method,heimowitz2020reducing}.
In particular, as we show next, averaging over all in-plane rotations using the PSWFs reduces the dimensionality of the problem, without any loss of information.
 We use autocorrelations up to and including the  third order. Indeed, second-order autocorrelations are not enough, as was observed already in~\cite{kam1980reconstruction} for a simpler problem where the input is not micrographs but rather picked, perfectly centered particles.

\subsection{Autocorrelation derivation} \label{sec:moment_derivation}

In this section, we prove relation~\eqref{eq:ac_micrographs}. We note that mathematically taking infinitely many micrographs is equivalent to taking one infinitely large micrograph with fixed density $\gamma$. Hence, we consider the moments of one micrograph $\II$ in the limit $N\to\infty$ and  $\gamma = \lim_{N\to\infty}\frac{MP^2}{N^2}\in(0,1)$. 
The separation condition guarantees that if $\mb i=(i,j)$ is in the support of some projection, then $\mb i + \mb \Bell$ for $\Bell\in[-(P-1),P-1]^2$ is either in the support of the same projection or outside the support of any projection. 

We begin by calculating the relation between the $p$th autocorrelation of the clean micrograph and the  averaged autocorrelation of the projections.
Let us denote the clean micrograph by $\widetilde \II = \II-\varepsilon$, where $\II$ and $\varepsilon$ are given in~\eqref{eq:micrograph_model}.     
Denote by $\mc S_t$ the support of the projection of the $t$-th particle in the micrograph. 
Then, we have
\begin{align}
	a^p_{\widetilde\II}[\Bell_1,\ldots, \Bell_{p-1}]  &= \frac{1}{N^2}\sum_{\mb i}\widetilde\II[\mb i]\widetilde\II[\mb i + \Bell_1]\cdots \widetilde\II[\mb i + \Bell_{p-1}]\nonumber\\
	&= \frac{1}{N^2}\sum_{t=1}^M\sum_{\mb i\in\mc S_t}\widetilde\II[\mb i]\widetilde\II[\mb i + \Bell_1]\cdots \widetilde\II[\mb i + \Bell_{p-1}] \nonumber\\ 
	&= \frac{MP^2}{N^2}\cdot\frac{1}{M}\sum_{t=1}^M\frac{1}{P^2}\sum_{i,j=0}^{P-1} I_{\omega_t}[\mb i]I_{\omega_t}[\mb i+\Bell_1]\cdots I_{\omega_t}[\mb i + \Bell_{p-1}]\nonumber\\
	&= \frac{MP^2}{N^2}\frac{1}{M}\sum_{t=1}^Ma^p_{I_{\omega_t}}[\Bell_1,\ldots, \Bell_{p-1}]\nonumber\\
	&\to \gamma\langle a^p_{I_{\omega}}[\Bell_1,\ldots, \Bell_{p-1}]\rangle_{\omega},
	\label{eq:general_moment}
\end{align}
where the average is taken over $\omega$ with respect to the distribution of viewing directions. Here, we assume it to be uniform.

In the presence of noise, we get additional bias terms denoted by $b_p$ in~\eqref{eq:ac_micrographs}. The mean ($p=1$) is unbiased,  since the noise is assumed to have zero mean. For the second-order autocorrelation ($p=2$), we have
\[\begin{aligned} 
	a^2_\II[\Bell] &=
	\frac{1}{N^2}\sum_{\mb i }\II[\mb i]\II[\mb
	i+\Bell]\\
	&= \frac{1}{N^2}\sum_{\mb i }\widetilde\II[\mb i]\widetilde\II[\mb i+\Bell] + \frac{1}{N^2}\sum_{\mb i }\widetilde\II[\mb i]\varepsilon[\mb i + \Bell]+ \frac{1}{N^2}\sum_{\mb i }\varepsilon[\mb i]\widetilde\II[\mb i + \Bell] + \frac{1}{N^2}\sum_{\mb i }\varepsilon[\mb i]\varepsilon[\mb i + \Bell]. 
\end{aligned}\]
The first term is given by~\eqref{eq:general_moment} for $p=2$. 
The cross terms have expected value $0$, and therefore almost surely vanish in the limit due to the law of large numbers.
The fourth term is zero unless $\Bell=0$, in which case it converges to $\sigma^2$.
Thus, we conclude
\begin{equation} \label{eq:second_order_ac_volume_data}
	a^2_\II[\Bell] \to \gamma\langle a^2_{I_{\omega}}[\Bell]\rangle_{\omega} + \sigma^2\delta[\Bell],
\end{equation}
where the  bias term $b_2[\Bell] = \sigma^2\delta[\Bell]$   depends only on the variance of the noise $\sigma^2$.

For the third moments, we get 8 terms:
\[\begin{aligned} 
	a^3_\II[\Bell_1, \Bell_2]  &
	= \underbrace{\frac{1}{N^2}\sum_{\mb i}\widetilde\II[\mb i]\widetilde\II[\mb i+\Bell_1]\widetilde\II[\mb i + \Bell_2]}_{(1)} +
	\underbrace{\frac{1}{N^2}\sum_{\mb i}\varepsilon[\mb i]\varepsilon[\mb i+\Bell_1]\varepsilon[\mb i + \Bell_2]}_{(2)}\\ 
	&+ \underbrace{\frac{1}{N^2}\sum_{\mb i}\widetilde\II[\mb i]\varepsilon[\mb i + \Bell_1]\widetilde\II[\mb i + \Bell_2]}_{(3)} +
	\underbrace{\frac{1}{N^2}\sum_{\mb i}\widetilde\II[\mb i]\widetilde\II[\mb i + \Bell_1]\varepsilon[\mb i + \Bell_2]}_{(4)}\\
	&+ \underbrace{\frac{1}{N^2}\sum_{\mb i}\varepsilon[\mb i]\widetilde\II[\mb i + \Bell_1]\widetilde\II[\mb i + \Bell_2]}_{(5)} +
	\underbrace{\frac{1}{N^2}\sum_{\mb i}\widetilde\II[\mb i]\varepsilon[\mb i + \Bell_1]\varepsilon[\mb i + \Bell_2]}_{(6)}\\
	&+ \underbrace{\frac{1}{N^2}\sum_{\mb i}\varepsilon[\mb i]\varepsilon[\mb i + \Bell_1]\widetilde\II[\mb i + \Bell_2]}_{(7)} +
	\underbrace{\frac{1}{N^2}\sum_{\mb i}\varepsilon[\mb i]\widetilde\II[\mb i + \Bell_1]\varepsilon[\mb i + \Bell_2]}_{(8)}.
\end{aligned}\]
We address these terms one by one:
\begin{itemize}
	\item  Term (1) is treated by~\eqref{eq:general_moment} for $p=3$;
	\item Term (2) is the third-order autocorrelation of  pure noise  which almost surely vanishes in the limit by the law of large numbers; 
	\item Terms (3)-(5) depend linearly on the noise and hence
	vanish in the limit;
	\item For term (6), if $\Bell_1\neq \Bell_2$ the term
	vanishes in the limit. If $\Bell_1 = \Bell_2$ then
	\[ \begin{aligned}
		\frac{1}{N^2}\sum_{\mb i}\widetilde\II[\mb i]\varepsilon[\mb i + \Bell]^2
		= \frac{MP^2}{N^2}\cdot\frac{1}{MP^2}\sum_{t=1}^{M}\sum_{\mb i\in\mc S_t}I_{\omega_t}[\mb i]\varepsilon[\mb i + \Bell]^2 
		\to \gamma\sigma^2 \langle a^1_{I_{\omega}}\rangle_{\omega}, 
	\end{aligned}\]
	where $\langle a^1_{I_{\omega}}\rangle_{\omega}$ is the mean of the volume. 
	\item  Terms (7) and (8) contribute $\delta$ functions similar to (6).
\end{itemize}
Thus, we conclude that
\begin{equation} \label{eq:third_order_ac_micro_volume}
	a^3_\II[\Bell_1, \Bell_2] \to \gamma\langle
	a^3_{I_{\omega}}[\Bell_1, \Bell_2]\rangle_{\omega}  +
	\gamma\sigma^2\langle a^1_{I_{\omega}}\rangle_{\omega}\Big(\delta[\Bell_1 - \Bell_2] +
	\delta[\Bell_1] + \delta[\Bell_2]\Big),
\end{equation}
where the second term is the bias $b_3[\Bell_1,\Bell_2]$.
Note that $\gamma\langle a^1_{I_{\omega}}\rangle_{\omega}$ is approximately the mean of the micrograph since  $a^1_\II \approx a^1_{\tilde\II} \approx \gamma\langle a^1_{I_{\omega}}\rangle_{\omega}$. Therefore, we do not need prior knowledge of $\gamma$ to effectively
debias the third-order autocorrelation.

\subsection{Accounting for all in-plane rotations} \label{sec:steering}

We represent our autocorrelations using Prolate Spheroidal Wave Functions (PSWFs)~\cite{slepian1964prolate}.
As we demonstrate below, this makes it easier to account for the fact that all in-plane rotations of the micrographs are equally likely observations. 
The PSWFs are given in polar coordinates by\footnote{A different normalization is used in \cite{landa2017steerable}.}
\begin{align}
	\psi_{k,q}(r,\varphi) & = \left\{\begin{array}{ll} \frac{1}{\sqrt{8\pi}}\alpha_{k,q}R_{k,q}(r)e^{\iota k\varphi}, & r\leq 1,\\ 0, & r>1,\end{array}\right. \label{eq:prolatesdef}
\end{align}
where $k,q$ are integers with $q\geq 0$, 
the range of $k,q$ is determined by Eq.~(8) in \cite{landa2017steerable}, the ${R_{k,q}}$ are a family of real, one-dimensional functions satisfying $R_{k,q}=R_{|k|,q}$, and the ${\alpha_{k,q}}$ are  scaling factors which will be defined in the next section.
The PSWFs are orthogonal on the unit disk.
Note that $\psi_{-k,q}=\overline{\psi_{k,q}}$, hence, we only need to consider PSWFs with $k\geq0$ when expanding real images.
 As can be seen from the definition of the  PSWFs~\eqref{eq:prolatesdef}, the effects of rotations and reflections on expansion coefficients of real images are, respectively, phase modulation and conjugation: this is why the PSWF basis is called \emph{steerable}~\cite{landa2017steerable,zhao2016fast}.
 Using the steerability property, we will show that the second-order autocorrelation, though two-dimensional, effectively only provides radial information.   
 Similarly, the 4-D third-order autocorrelation truly only carries information along three dimensions.
While there exist alternative steerable bases, such as Fourier-Bessel~\cite{zhao2016fast}, we chose to work with the PSWFs since they are eigenfunctions of the truncated Fourier transform: a fact that we exploit later on.

We start with the second-order autocorrelation.
For $\Bell \in [-(P-1), P-1]^2$, let us define 
\begin{equation} \label{eq:PSWF_expansion}
	\mathfrak{a}^2[k, q]  =
	\sum_{\Bell}a^2_\II[\Bell]\overline{\psi_{k,q}[\Bell]}, 
\end{equation}
where $\psi_{k,q}[\Bell] := \psi_{k,q}(\Bell/(P-1))$ is a discretization of the PSWFs. 
Knowledge of these coefficients is essentially equivalent to knowledge of the second-order correlations owing to the following approximate identity:
\begin{align}
	a^2_\II[\Bell] & \approx \sum_{k,q}\mathfrak{a}^2[k,q]\psi_{k,q}[\Bell].
	\label{eq:approxprolatesdiscrete}
\end{align}
This holds because the continuous PSWFs form an orthonormal basis, and their discretized counterparts with indices $(k,q)$ appropriately bounded as in~\cite{landa2017steerable} are (empirically) almost orthonormal.  
As a result, for our purposes, the pair of equations above provides a basis expansion for the autocorrelations.

We now proceed to show that the coefficients $\mathfrak{a}^2[k, q]$ can be computed from the micrographs directly.
By definition,
\begin{align}
	\mathfrak{a}^2[k,q] & =\sum_{\Bell} a^2_\II[\Bell] \overline{\psi_{k,q}[\Bell]} \nonumber\\
	& = \frac{1}{N^2}\sum_{\mb i}\II[\mb i]\left(\sum_{\Bell}\II[\mb i+\Bell]\overline{\psi_{k,q}[\Bell]}\right) \nonumber \\
	& = \frac{1}{N^2}\sum_{\mb i}\II[\mb i] a_{k,q}[\mb i], \label{eq:frakatwo}
\end{align}
where we defined,
\begin{align}
	a_{k,q}[\mb i] & = \sum_{\Bell}\II[\mb i+\Bell]\overline{\psi_{k,q}[\Bell]}.
	\label{eq:patchcoefficients}
\end{align}
These coefficients can be computed efficiently. Indeed, consider a patch of the micrograph $\II$ centered around pixel $\mb i$ and of size $(2P-1) \times (2P-1)$. This is exactly the patch indexed in the sum above. Hence, using the same approximation as we did in~\eqref{eq:approxprolatesdiscrete}, a direct expansion of that patch in the discretized PSWFs yields the sought coefficients:
\begin{align}
	\II[\mb i+\Bell] & \approx \sum_{k,q} a_{k,q}[\mb i] \psi_{k,q}[\Bell]. \label{eq:itamar}
\end{align}
Thus, we proceed as follows:
for each position $\mb i$ in the micrograph $\II$, we extract the corresponding patch of size $(2P-1) \times (2P-1)$, expand it in the discretized PSWFs as in~\eqref{eq:itamar}, and collect the $a_{k,q}$ as per~\eqref{eq:frakatwo} to constitute the second-order autocorrelation of the micrograph.

Following this formalism, and using the steerability property, it is now straightforward to account for all in-plane rotations and reflections of the micrograph. We have the following approximate expansions for a patch rotated about its center $\mb i$ by an angle $\alpha$:
\[ \II^{\alpha,+}[\mb i+\Bell] \approx \sum_{k,q}a_{k,q}e^{-\iota k\alpha}\psi_{k,q}[\Bell],\]
and the reflection followed by a rotation by angle $\alpha$:
\[ \II^{\alpha,-}[\mb i+\Bell] \approx \sum_{k,q}\overline{a_{k,q}}e^{-\iota k\alpha}\psi_{k,q}[\Bell].\]
Averaging over all rotations of the patch $\II(\mb i + \Delta\mb i)$ and its reflection we get
\begin{align} \label{eq:second_order_ac_pswf}
	\mathfrak{a}^2[k,q] &= \frac{1}{N^2}\sum_{\mb i}\II[\mb i]\left(\frac{1}{4\pi}\int_0^{2\pi}\left(a_{k,q}[\mb i] +
	\overline{a_{k,q}}[\mb i]\right)e^{-\iota k\alpha}\, d\alpha\right)\nonumber\\ 
	&= \delta[k] \frac{1}{N^2}\sum_{\mb i}\II[\mb i]a_{0,q}[\mb i], 
\end{align}
where in the last equality we used that $a_{0,q}[\mb i]$ is real since both $\II$ and $\psi_{0,q}$ are real valued (more generally, $a_{-k,q}=\overline{a_{k,q}}$).
Thus, the second-order autocorrelation, though two-dimensional, effectively only provides radial information.

We now follow a similar approach to estimate the bias term $b_2$.
Introduce the coefficients $\mathfrak{b}_2$ as:
\[\begin{aligned} 
	\mathfrak{b}_2[k,q] & = \sigma^2\sum_{\Bell}\delta[\Bell]\overline{\psi_{k,q}}[\Bell]\\
	&= \sigma^2\overline{\psi_{k,q}}[\mb 0]\\
	&= \delta[k] \frac{\sigma^2}{\sqrt{2\pi}}R_{0,q}(0), \end{aligned}\]
where we used the fact that the functions $R_{k,q}$ are zero at the origin for $k\neq 0$.
With this definition, we have the usual approximation:
\[ b_2[\Bell] = \sigma^2\delta[\Bell] \approx \sum_{k,q}\mathfrak{b}_2[k,q]\psi_{k,q}[\Bell].\]

We now turn our attention to the third-order autocorrelation. Following the same lines, we define the coefficients:
	\[\begin{aligned} \mathfrak{a}^3[k_1,q_1;k_2,q_2] &= \sum_{\Bell_1, \Bell_2} a^3_\II[\Bell_1,\Bell_2]\overline{\psi_{k_1,q_1}[\Bell_1]}\psi_{k_2,q_2}[\Bell_2]\\
		&= \frac{1}{N^2}\sum_{\mb i}\II[\mb i]\left(\sum_{\Bell_1}\II[\mb i+\Bell_1]\overline{\psi_{k_1,q_1}[\Bell_1]}\right) \left(\sum_{\Bell_2}\II[\mb i+\Bell_2]\psi_{k_2,q_2}[\Bell_2]\right)\\
		&= \frac{1}{N^2}\sum_{\mb i}\II[\mb i] a_{k_1,q_1}[\mb i]\overline{a_{k_2,q_2}[\mb i]},
	\end{aligned}\]
	where the patch expansion coefficients $a_{k,q}$ are as defined in~\eqref{eq:patchcoefficients}.
	The coefficients $\mathfrak{a}^3[k_1,q_1;k_2,q_2]$ are related to the third-order autocorrelation via the approximate identity:
	\[ a^3_\II[\Bell_1, \Bell_2] \approx \sum_{\substack{k_1,q_1\\ k_2,q_2}}\mathfrak{a}^3[k_1,q_1;k_2,q_2]\psi_{k_1,q_1}[\Bell_1]\overline{\psi_{k_2,q_2}[\Bell_2]}.\]
	Averaging over all rotations of $\II$ and its reflection, we obtain
	\begin{align} \label{eq:third_order_ac_pswf}
		\mathfrak{a}^3[k_1,q_1;k_2,q_2] &= \frac{1}{N^2}\sum_{\mb i}\II[\mb i] \frac{1}{4\pi}\int_0^{2\pi}\left(a_{k_1,q_1}[\mb i]\overline{a_{k_2,q_2}[\mb i]} + \overline{a_{k_1,q_1}[\mb i]}a_{k_2,q_2}[\mb i]\right)e^{-\iota(k_1-k_2)\alpha}d\alpha \nonumber\\
		&= \delta[k_1 - k_2]\frac{1}{N^2}\sum_{\mb i}\II[\mb i]\Re\{a_{k_1,q_1}[\mb i]\overline{a_{k_2,q_2}[\mb i]}\}. 
	\end{align}
	Thus, similarly to the second-order autocorrelation, averaging over all in-plane rotations reveals that the 4-D third-order autocorrelation truly only carries information along three dimensions.

	Finally, we treat the bias terms:
	\begin{align*}
		\mk b_3[k_1,q_1; k_2,q_2] = \gamma\sigma^2\langle a^1_{I_{\omega}} \rangle_\omega \delta[k_1 - k_2]\left[\delta[q_1-q_2] + \delta[k_1]\frac{1}{2\pi}(\alpha_{0,q_1} + \alpha_{0,q_2})R_{0,q_1}(0)R_{0,q_2}(0)\right].
	\end{align*}
	Thus,
	\begin{align*}
		b_3[\Bell_1, \Bell_2] & = \gamma\sigma^2\langle a^1_{I_{\omega}}\rangle_{\omega}\Big(\delta[\Bell_1 - \Bell_2] +
		\delta[\Bell_1] + \delta[\Bell_2]\Big) \\
		& \approx \sum_{\substack{k_1, q_1\\ k_2, q_2}} \mk 
		b_3[k_1,q_1; k_2, q_2]\psi_{k_1,q_1}[\Bell_1]\overline{\psi_{k_2,q_2}[\Bell_2]}.
	\end{align*}
	
	\subsection{Connection to volume}

	Until now, we have established simple relations between the autocorrelations of the micrographs and the autocorrelations of the projection images of the 3-D structure, and explained how to compute the former.  
	Now, we complete the picture by relating the latter to the 3-D structure itself.
	
	Using the 2-D PSWFs and the Fourier slice theorem, we can express each projection~\eqref{eq:projection_model} as 
	\[ \widehat I_{\omega}(ck,\theta) = \sum_{N,n}b_{N,n}(\omega)\psi_{N,n}(k,\theta),\]
	where
		\begin{align*}
			b_{N,n}(\omega) &= \frac{4}{\sqrt{2\pi}|\alpha_{N,n}|^2}\int_0^{2\pi}\int_0^1\widehat I_{\omega}(ck,\theta)R_{N,n}(k)e^{-\iota N\theta}k\, dk\, d\theta,\\
			&=\sum_{\ell,m,m',s}\tamir_{\ell,m,s}\left[\frac{\sqrt{8\pi}}{\alpha_{N,n}}Y_{\ell}^{m'}(\pi/2,0)\right]D_{m',m}^{\ell}(\omega) \left(\int_0^1j_{\ell,s}(k)R_{N,n}(k)k\, dk\right)\left(\frac{1}{2\pi}\int_0^{2\pi}e^{\iota(m'-N)\theta}\, d\theta\right)\\
			&= \sum_{\vert N\vert\leq \ell}\sum_{m,s}\tamir_{\ell,m,s}D_{N,m}^{\ell}(\omega)\beta_{\ell,s;N,n},
		\end{align*}
		and the coefficients
		\begin{equation} \label{eq:beta}
			\beta_{\ell,s;N,n} := \left\{\begin{array}{ll} \frac{\sqrt{8\pi}}{\alpha_{N,n}}Y_{\ell}^{N}(\pi/2,0)\int_0^1j_{\ell,s}(k)R_{N,n}(k)k\, dk, & \ell\geq |N|\\ 0, & \ell<|N|\end{array}\right.,
		\end{equation}
		can be precomputed.
	The PSWFs are eigenfunctions of the truncated Fourier transform~\cite{landa2017steerable} and hence satisfy 
		\begin{equation}
			\label{eq:PSWF_defn_eq}
			\alpha_{N,n}\psi_{N,n}(\mb k) = \int\limits_{||\mb r||_2\leq 1} \psi_{N,n}(\mb r)e^{\iota c(\mb r\cdot\mb k)}d\mb r.
		\end{equation} 
		We can now express the projection in real space as
		\begin{align} \label{eq:proj_expansion}
			&I_{\omega}(r,\varphi) = \sum_{N,n}\widehat{\alpha}_{N,n}b_{N,n}(\omega)\psi_{N,n}(r,\varphi)\\
			&= \sum_{\ell=0}^{L_{\text{max}}}\sum_{N,m=-\ell}^{\ell}\sum_{n=0}^{n_{\text{max}}(N)}\sum_{s=1}^{S(\ell)}\tamir_{\ell,m,s}\widehat\beta_{\ell,s;N,n}D_{N,m}^{\ell}(\omega)\psi_{N,n}(r,\varphi), \nonumber 
		\end{align}
		where $n_{\text{max}}(N)$ is chosen according to Eq.~(8) in \cite{landa2017steerable}, $\alpha_{N,n}$ is the eigenvalue corresponding to the $(N,n)$th PSWF, $\widehat{\alpha}_{N,n} = (c/2\pi)^2\alpha_{N,n}$, and $\widehat\beta_{\ell,s;N,n}=\widehat\alpha_{N,n}\beta_{\ell,s;N,n}$.

		\subsubsection{First-order autocorrelation (the mean)}
		
		Since $j_{\ell,s}(0) = 0$ unless $\ell=0$, and since
		$Y_{0,0}(\theta,\varphi) = \frac{1}{\sqrt{4\pi}}$, we conclude from~\eqref{eq:volume_expansion_app} that
		\begin{equation} \label{eq:ac_1_prolates}
			\mathfrak{a}_x^1 = \langle
			a^1_{I_{\omega}}\rangle_{\omega}  = \widehat  \phi(\mb 0) = \frac{1}{\sqrt{4\pi}}\sum_s\tamir_{0,0,s}j_{0,s}(0).
		\end{equation}
		
		\subsubsection{Second-order autocorrelation}
		The second-order autocorrelation is easier to derive directly in Fourier space, to avoid integration of shifted PSWFs against centered ones. 
		The relation between the second-order autocorrelation of the micrographs and projection images of the volume is given in~\eqref{eq:second_order_ac_volume_data} and~\eqref{eq:second_order_ac_pswf}. 
		The connection with the expansion coefficients of the volume can be derived in  Fourier space directly from Kam's
		original formula~\cite[Eq. (11)]{kam1980reconstruction} by setting $\mb k_1 = \mb k_2$ to
		obtain,
		\begin{align*}
			\left\langle a^2_{\hat{I}_{\omega}}(k,\theta)\right\rangle_{\omega} &=
			\frac{1}{4\pi}\sum_{\ell,
				m}\left|\sum_s\tamir_{\ell,m,s}j_{\ell,s}(k)\right|^2 \\&=
			\frac{1}{4\pi}\sum_{\substack{\ell,m\\s_1,s_2}}\tamir_{\ell,m,s_1}\overline{\tamir_{\ell,m,s_2}}j_{\ell,s_1}(k)j_{\ell,s_2}(k),
		\end{align*}
		where we used the fact that the normalized spherical Bessel functions
		$j_{\ell,s}$ are real. 
		
		As before, we want to derive the relation with respect to the PSWF coefficients of the autocorrelation. Hence,  we expand the above in 2-D PSWFs by
		\[\left\langle a^2_{\hat{I}_{\omega}}(k,\theta)\right\rangle_{\omega} =
		\sum_{q}\mathfrak{a}_x^2(q)\psi_{0,q}(k),\]
		and conclude that 
		\begin{equation} \label{eq:ac_2_prolates}
			\mathfrak{a}_x^2[q] =
			\frac{1}{\sqrt{8\pi}}\sum_{\substack{\ell,m\\s_1,s_2}}\tamir_{\ell,m,s_1}\overline{\tamir_{\ell,m,s_2}}
			\int_0^1j_{\ell,s_1}(k)j_{\ell,s_2}(k)R_{0,q}(k)k\, dk.
		\end{equation}
		The integral on $k$ is precomputed.

		\subsubsection{Third-order autocorrelation}
		In~\eqref{eq:third_order_ac_micro_volume} and~\eqref{eq:third_order_ac_pswf} we have shown how the third-order autocorrelations of the micrographs and the projection images of the volume are related, and how we can represent them in PSWFs. Now, we relate these expressions to the expansion coefficients of the volume itself.
		
		The third-order autocorrelation of the volume can be expressed in terms of~\eqref{eq:proj_expansion}: 
		\begin{align} \label{eq:third_order_ac_volume}
			\langle
			a^3_{I_{\omega}}[\Bell_1, \Bell_2]\rangle_{\omega} &= \frac{1}{L^2}\sum_{\mb i}\langle I_{\omega}[\mb i]I_{\omega}[\mb i+\Bell_1]\overline{I_{\omega}([\mb i+\Bell_2])}\rangle_{\omega},\\
			&\approx \frac{1}{L^2}\sum_{\substack{N_1,n_1\\N_2,n_2\\N_3,n_3}} \langle b_{N_1,n_1}(\omega)b_{N_2,n_2}(\omega)\overline{b_{N_3,n_3}}(\omega)\rangle_{\omega} \sum_{\mb i}\psi_{N_1,n_1}[\mb i]\psi_{N_2,n_2}[\mb i+\Bell_1]\overline{\psi_{N_3,n_3}([\mb i+\Bell_2])},\nonumber
		\end{align}
		where the approximation is due to discretization. 
		Now,
			\[\begin{aligned} \langle b_{N_1,n_1}(\omega)b_{N_2,n_2}(\omega)\overline{b_{N_3,n_3}}(\omega)\rangle_{\omega} &= \sum_{\substack{\ell_1,m_1,s_1\\\ell_2,m_2,s_2\\\ell_3,m_3,s_3}}\tamir_{\ell_1,m_1,s_1}\tamir_{\ell_2,m_2,s_2}\overline{\tamir_{\ell_3,m_3,s_3}}\\&\times \langle D_{N_1,m_1}^{\ell_1}(\omega)D_{N_2,m_2}^{\ell_2}(\omega)\overline{D_{N_3,m_3}^{\ell_3}}(\omega)\rangle_{\omega} \widehat\beta_{\ell_1,s_1;N_1,n_1}\widehat\beta_{\ell_2,s_2;N_2,n_2}\overline{\widehat\beta_{\ell_3,s_3;N_3,n_3}},\end{aligned}\]
			where the latter coefficients are given explicitly in~\eqref{eq:beta}. Using standard properties of Wigner-D functions, we obtain  
			\begin{align*}
				\left\langle D_{N_1,m_1}^{\ell_1}(\omega)D_{N_2,m_2}^{\ell_2}(\omega)\overline{D_{N_3,m_3}^{\ell_3}}(\omega)\right\rangle_{\omega} = (-1)^{N_3+m_3} \left(\begin{array}{ccc}\ell_1 & \ell_2  & \ell_3\\ N_1 & N_2 & -N_3\end{array}\right)\left(\begin{array}{ccc}\ell_1 & \ell_2  & \ell_3\\ m_1 & m_2 & -m_3\end{array}\right),
			\end{align*}
			where  $\left(\begin{array}{ccc} \ell_1 & \ell_2 & \ell_3\\ m_1 & m_2 & m_3\end{array}\right)$ are called Wigner 3-j symbols. 
			Notably, these terms are zero unless $m_1+m_2+m_3=0$ and $|\ell_1-\ell_2|\leq \ell_3\leq \ell_1+\ell_2$. Thus,  we conclude that 
			\begin{align} \label{eq:b_average}
				\langle b_{N_1,n_1}(\omega)b_{N_2,n_2}(\omega)\overline{b_{N_3,n_3}}(\omega)\rangle_{\omega} &= \delta_{N_3,N_1+N_2} \sum_{\substack{\ell_1,m_1,s_1\\\ell_2,m_2,s_2\\s_3}}\sum_{\ell_3=|\ell_1-\ell_2|}^{\min(L,\ell_1+\ell_2)}\tamir_{\ell_1,m_1,s_1}\tamir_{\ell_2,m_2,s_2}\overline{\tamir_{\ell_3,m_1+m2,s_3}}\\
				&\times (-1)^{N_1+N_2+m_1+m_2}\left(\begin{array}{ccc}\ell_1 & \ell_2  & \ell_3\\ N_1 & N_2 & -N_1-N_2\end{array}\right)\nonumber\\&\times \left(\begin{array}{ccc}\ell_1 & \ell_2  & \ell_3\\ m_1 & m_2 & -m_1-m_2\end{array}\right) \widehat\beta_{\ell_1,s_1;N_1,n_1}\widehat\beta_{\ell_2,s_2;N_2,n_2}\overline{\widehat\beta_{\ell_3,s_3;N_1+N_2,n_3}}. \nonumber
			\end{align}
			Combining~\eqref{eq:b_average} with~\eqref{eq:third_order_ac_volume} provides the explicit relation between the third-order autocorrelation and the volume.   
			
			Recall that we obtain the autocorrelations of the volume in PSWFs coefficients; see~\eqref{eq:third_order_ac_pswf}. Hence, to conclude the derivation we expand
			\[ \langle a^3_{I_{\omega}}[\Bell_1, \Bell_2]\rangle_{\omega}= \sum_{k,q_1,q_2}\mathfrak{a}_x^3[k,q_1,q_2]\psi_{k,q_1}[\Bell_1]\overline{\psi_{k,q_2}[\Bell_2]},\]
			where we only include the block-diagonal terms in the expansion; the rest are equal to zero. Let,
			\begin{align*}
				\Psi_{\ell,N,s}[\Bell] &:= \sum_{n=0}^{n_{\text{max}}(N)}\widehat\beta_{\ell,s;N,n}\psi_{N,n}[\Bell],\\ 
				\rho_{\ell,N,s}^{(k,q)}&:=\int_{\Bell}\Psi_{\ell,N,s}([\mb i+\Bell])\overline{\psi_{k,q}(\Bell)}.
			\end{align*}
			Then, 
			the final formula reads
			\begin{align} \label{eq:ac_3_prolates}
				\mathfrak{a}_x^3[k,q_1,q_2] &= \sum_{\substack{\ell_1,m_1,s_1\nonumber \\ \ell_2,m_2,s_2\\s_3}}\sum_{\ell_3=|\ell_1-\ell_2|}^{\min(L,\ell_1+\ell_2)}\tamir_{\ell_1,m_1,s_1}\tamir_{\ell_2,m_2,s_2}\overline{\tamir_{\ell_3,m_1+m_2,s_3}}\\
				&\times (-1)^{m_1+m_2}\left(\begin{array}{ccc}\ell_1 & \ell_2  & \ell_3\\ m_1 & m_2 & -m_1-m_2\end{array}\right)\\
				&\times \sum_{N_1=-\ell_1}^{\ell_1}\sum_{N_2=-\ell_2}^{\ell_2}(-1)^{N_1+N_2}\left(\begin{array}{ccc}\ell_1 & \ell_2  & \ell_3\nonumber\\ N_1 & N_2 & -N_1-N2\end{array}\right)\nonumber\\&\times\frac{1}{L^2}\sum_{\mb i}\Psi_{\ell_1,N_1,s_1}[\mb i]\rho_{\ell_2,N_2,s_2}^{(k,q_1)}[\mb i]\overline{\rho_{\ell_3,N_1+N_2,s_3}^{(k,q_2)}[\mb i]}. \nonumber
			\end{align}
			In practice, the last two lines of the above expression for $\mathfrak{a}_x^3[k,q_1,q_2]$ are precomputed, and both the integration over $\mb i$ and over $\Bell$ are performed on the grid of the images in the data set, to match the integration performed on the actual images.

\subsection{Recovering the volume from the autocorrelations of the micrographs}

To estimate the coefficients of the volume itself, we solve the least squares problem
\begin{equation} \label{eq:LS_cryo}
	\min_{\tilde{x},\tilde{\gamma}} w_1 \vert \mathfrak{a}_x^1 - \hat{\gamma}\mathfrak{a}_{\tilde{x}}^1 \vert^2 + w_2\| \mathfrak{a}_x^2 - \tilde{\gamma}\mathfrak{a}_{\tilde{x}}^2 \|_2^2 + w_3\| \mathfrak{a}_x^3 - \tilde{\gamma}\mathfrak{a}_{\tilde{x}}^3 \|_\text{F}^2, 
\end{equation}  
where the explicit expressions of $\mathfrak{a}_x^1$, $\mathfrak{a}_x^2$ and $\mathfrak{a}_x^3$ are given in~\eqref{eq:ac_1_prolates},~\eqref{eq:ac_2_prolates} and~\eqref{eq:ac_3_prolates}, respectively.  
In the experiments, we set $w_1=w_2=w_3=1$. 
We solve the least squares problem using Matlab's \texttt{lsqnonlin} solver for nonlinear least squares problems with the trust-regions algorithm.
The algorithm is initialized by sampling the coefficients from i.i.d.\ random Gaussians.

\subsection{Complexity analysis}

The computational complexity for the computation of the moments from the micrograph is $O(KN^2P^6)$, since we have $N^2$ patches that we extract from each of $K$ micrographs. For each patch, we perform an expansion in PSWFs with complexity $O(P^3)$~\cite{landa2017steerable}, and then compute the autocorrelations dominated by computation of the third-order autocorrelation as in \eqref{eq:third_order_ac_pswf} costing $O(P^6)$. 
We note that computing autocorrelations can be executed efficiently on CPUs and GPUs, and in parallel across micrographs. It can even be done in a  streaming mode, as only one pass through each micrograph is necessary. 
The complexity of evaluating the third-order autocorrelation using~\eqref{eq:ac_3_prolates} after the precomputation takes $O(BP^6)$, where $B$ is the number of entries in the third-order autocorrelation. Since $B=O(P^3)$, the total complexity for evaluating the third-order autocorrelation is $O(P^9)$. That is because, if the number of volume expansion coefficients is $V = O(P^3)$, the dominant step can be written as a matrix-vector multiplication with a matrix of size $BV\times V$ and a vector $V\times 1$, so the cost is $BV^2=O(P^9)$. 
The complexity of solving the least squares problem~\eqref{eq:LS_cryo} is $O(TP^9)$, where $T$ is the number of iterations required for the optimizer to converge, since the third-order autocorrelation evaluation dominates the cost of each iteration. The optimizer terminates when the norm of the gradient of the cost function in~\eqref{eq:LS_cryo} decreases below $10^{-6}$ or the number of iterations exceeds $10^4$. 

\section{Numerical experiments} \label{sec:numerical_experiments}

The technique we advocate allows recovery of a 3-D structure from its tomographic projections hidden in noisy micrographs without detecting their locations. To illustrate the underlying principles of the method, we present several simple proof-of-concept results for simulated cryo-EM data.\footnote{The code to generate all figures is publicly available in \url{https://github.com/PrincetonUniversity/BreakingDetectionLimit}.}
We first present  numerical results, and then provide additional technical details.

\begin{figure}[h]
	\centering
	\begin{subfigure}[h]{0.45\textwidth}
		\centering
		\includegraphics[angle=270, width=1\textwidth]{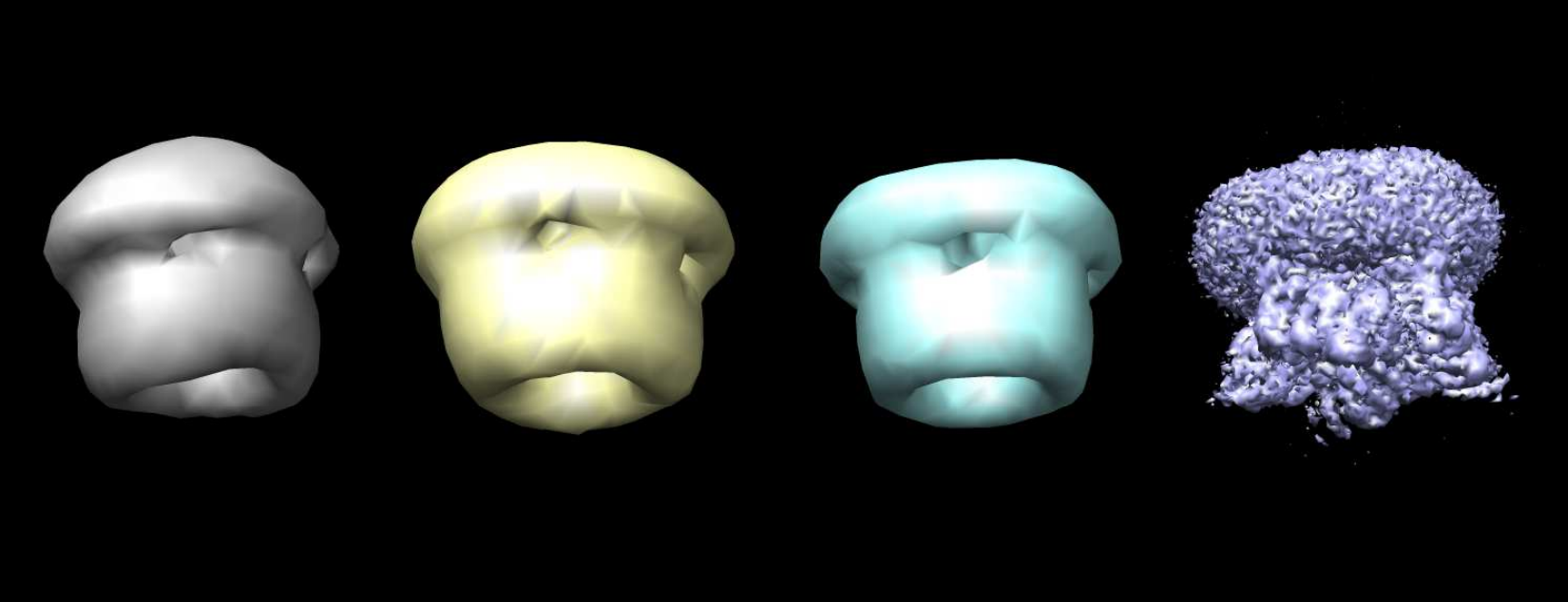}
		\caption{TPRV1 with cutoff $L_{\text{max}}=5$}\label{fig:TRPV1_recon}
	\end{subfigure}
	\hfill
	\begin{subfigure}[h]{0.45\textwidth}
		\centering
		\includegraphics[angle=270, width=1\textwidth]{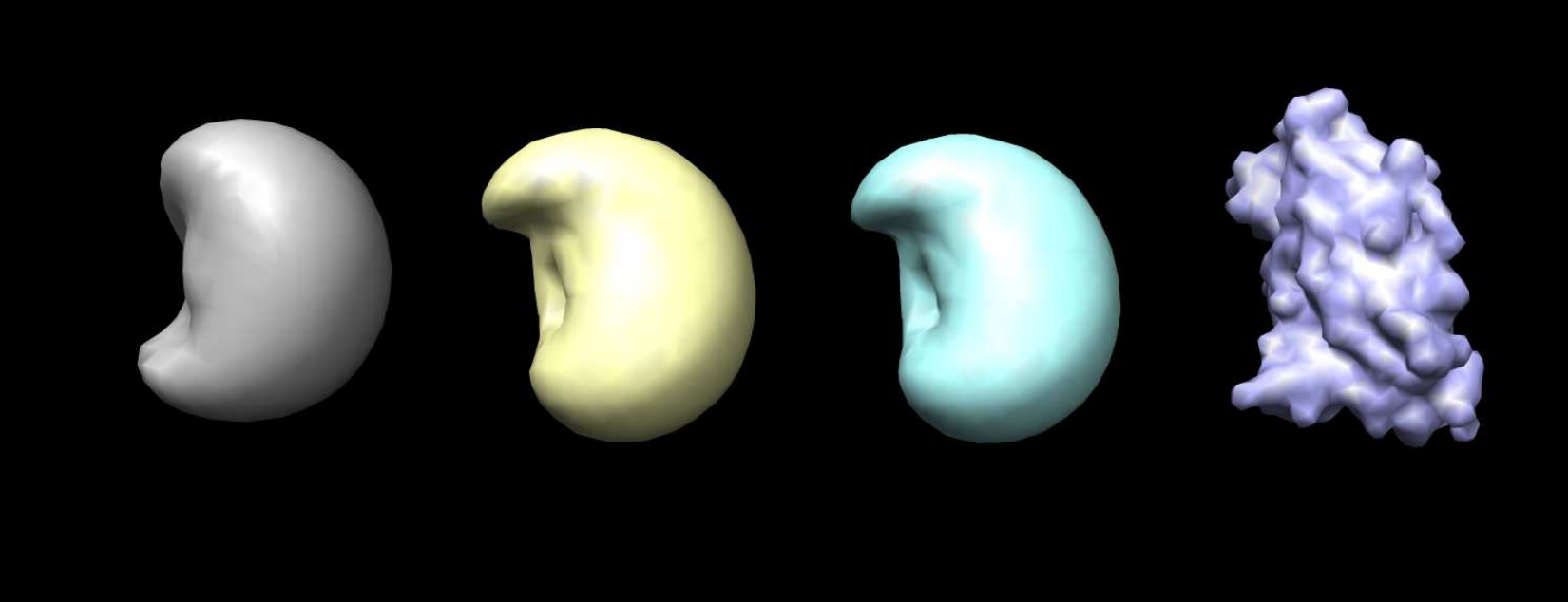}
		\caption{BPTI with cutoff $L_{\text{max}}=2$}\label{fig:BPTI_recon}
	\end{subfigure} 
	\caption{\label{fig:cryo_recon} Reconstructions from the first, second and third order autocorrelations.
		The ground truth volumes were expanded according to~\eqref{eq:volume_expansion_app} with cutoff $L_{\text{max}}$.
		{The original molecules are shown in purple and the (smoothed) ground truths in blue  to illustrate the smoothing effect of our downsampling and truncation of the spherical harmonics expansion.
			The reconstructions  from clean autocorrelations are shown in yellow, and recoveries from 
			autocorrelations estimated from noisy data in gray. For the noisy experiments, we used 300 micrographs with SNRs of 1/16 for TRPV1 and 1/64 for BPTI. We present excerpts from noisy micrographs side-by-side with the corresponding clean ones in Figure~\ref{fig:cryo_noisy_micros_for_rec}}.}
\end{figure}

\begin{figure}[t!]
	\centering
	\begin{subfigure}[t]{0.245\textwidth}
		\centering
		\includegraphics[scale=0.3]{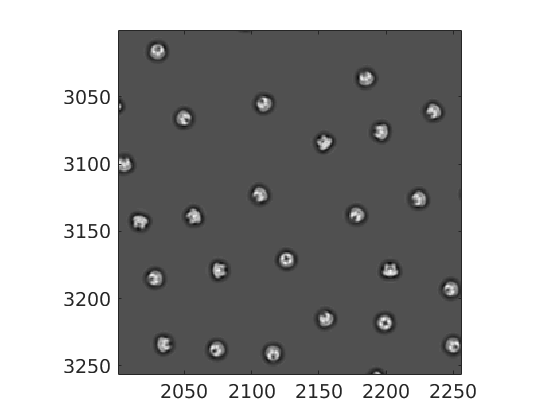}
		\caption{}
	\end{subfigure} \hfill 
	\begin{subfigure}[t]{0.245\textwidth}
		\centering
		\includegraphics[scale=0.3]{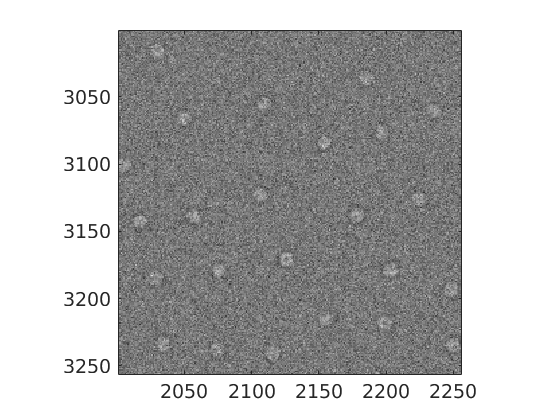}
		\caption{}
	\end{subfigure} \hfill 
	\begin{subfigure}[t]{0.245\textwidth}
		\centering
		\includegraphics[scale=0.3]{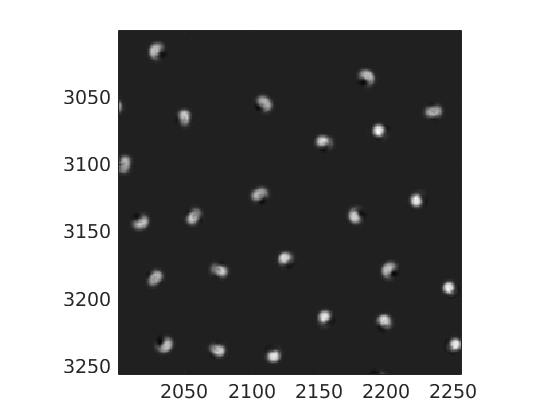}
		\caption{}
	\end{subfigure} \hfill
	\begin{subfigure}[t]{0.245\textwidth}
		\centering
		\includegraphics[scale=0.3]{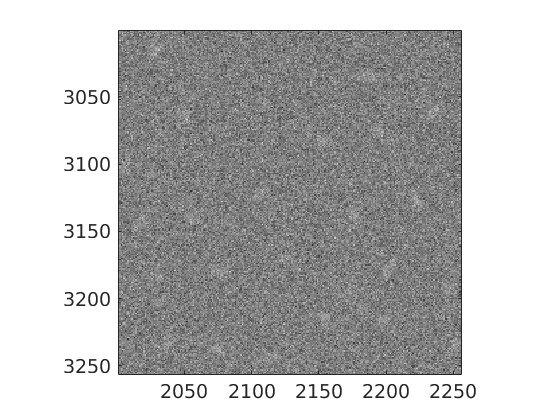}
		\caption{}
	\end{subfigure}
	\caption{\label{fig:cryo_noisy_micros_for_rec}  {Excerpts of size $250\times 250$ from the noisy micrographs used for the reconstructions in Figure~\ref{fig:cryo_recon} and the corresponding clean excerpts. (a) Excerpt of the clean TRPV1 micrograph; (b) Excerpt of the noisy TRPV1 micrograph with SNR $= 1/16$; (c) Excerpt from the clean BPTI micrograph; (d) Excerpt from the noisy BPTI micrograph with SNR $= 1/64$.}} 
\end{figure}

\begin{figure}[t!]
	\centering
	\begin{subfigure}[t]{0.4\textwidth}
		\centering
		\includegraphics[scale=0.35]{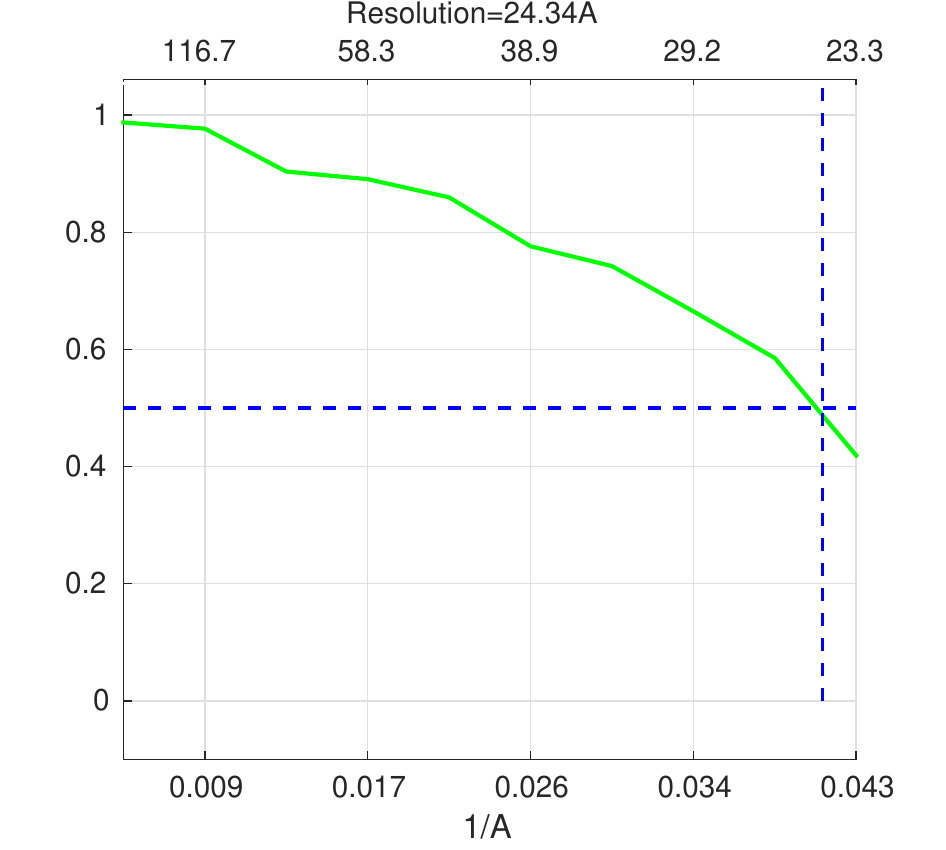}
		\caption{}
	\end{subfigure} \hfill 
	\begin{subfigure}[t]{0.4\textwidth}
		\centering
		\includegraphics[scale=0.35]{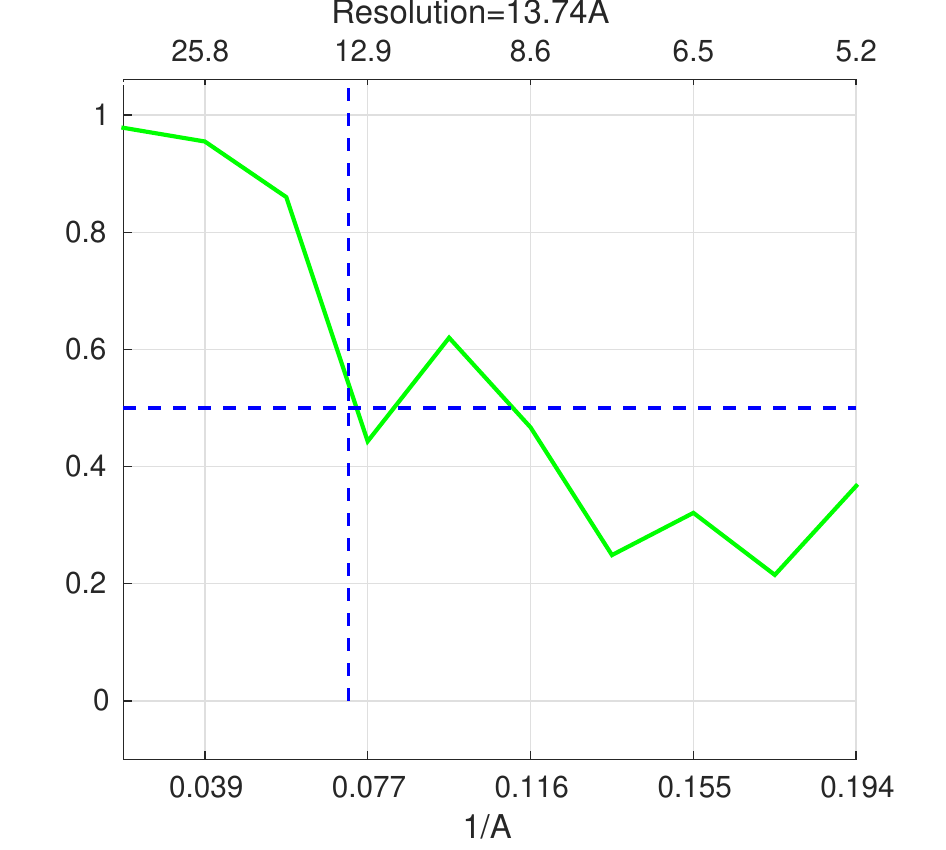}
		\caption{}
	\end{subfigure} 
	\caption{\label{fig:cryo_FSC} {Fourier Shell Correlations (FSCs) for the noisy reconstructions presented in Figure~\ref{fig:cryo_recon}.
The FSC measures the normalized cross-correlation  between  the recovered 3-D structure and the (smoothed) ground truth 3-D structure  over corresponding shells in Fourier space (i.e., as a function of spatial frequency). The resolution is determined as the frequency where the FSC curve drops below $0.5$. This is the standard resolution measure in the cryo-EM literature.
			(a) FSC for the TRPV1 reconstruction, giving resolution of $24$\AA; (b) FSC for the BPTI reconstruction, giving resolution of $13$\AA.}} 
\end{figure}

\paragraph{Numerical results.}
 Figure~\ref{fig:cryo_recon} shows recoveries of the 3-D volumes from the clean autocorrelations and from 300  noisy micrographs of size $7420^2$. 
 The experiments were conducted on  the Bovine Pancreatic Trypsin Inhibitor (BPTI) and the TRPV1 molecules (see technical details at the end of this section).
In the experiments, we define SNR  as $\text{SNR} = \frac{\text{var}(\mathcal{I})}{\sigma^2}$, where $\text{var}(\mathcal{I})$ is the variance of our stack of micrographs and $\sigma^2$ is the variance of noise. The noise level in Figure~\ref{fig:cryo_noisy_micros_for_rec} was $\text{SNR} = 1/16$ for the TRPV1 micrographs and $\text{SNR} = 1/64$ for the BPTI micrographs. 
We present excerpts from noisy micrographs side-by-side with the corresponding clean ones in Figure~\ref{fig:cryo_noisy_micros_for_rec}.
Figure~\ref{fig:cryo_FSC} presents the Fourier shell correlation curves. 

Numerical evidence suggests that autocorrelations up to order three, together, uniquely determine the 3-D volume (see also~\cite{bandeira2017estimation,fan2021maximum} for statistical analysis of  closely-related models). 
Unfortunately, the mapping between the autocorrelations and volume seems to be ill-conditioned, preventing high-resolution recovery from noisy data.
For the TRPV1 reconstruction, the optimizer converged to an estimator with relative $\ell_2$ error of $10^{-6}$ in the first three autocorrelations and an error of $10^{-1}$ in the expansion coefficients of the volume. For the BPTI reconstruction, the errors in the autocorrelations were $10^{-6}$, $10^{-7}$ and $10^{-8}$ for the third, second and first autocorrelations, respectively, while the error in the expansion coefficients of the volume was $5\times 10^{-2}$. This illustrates the ill-conditioning of the map between the volume and its first three autocorrelations that prevents us from obtaining {high-resolution} results from noisy data.
In the next section, we outline how we suggest overcoming the ill-conditioning in future work.

While we cannot provide a {high-resolution} 3-D reconstruction from noisy data with the current algorithm, our method can be easily applied to the problem of deciding whether a micrograph contains projections or merely pure noise---a problem considered in classical works in statistics~\cite{donoho2004higher} 
and cryo-EM~\cite{henderson2013avoiding}.
This task can be performed by considering solely the recovered $\gamma$ (the fraction of pixels occupied by projections in the micrograph), estimated as part of the recovery algorithm.
Specifically, for this experiment, 
we used 25 micrographs of size $7420^2$, and
the LS problem~\eqref{eq:LS_cryo} was solved assuming the spherical harmonic cutoff for the volume is $L_{\text{max}}=0$, which is sufficient to recover a significant $\widehat\gamma$ in the presence of projections in the micrograph.

\begin{figure}[h]
	\centering
	\begin{subfigure}[t]{0.3\textwidth}
		\centering
		\includegraphics[scale=0.4]{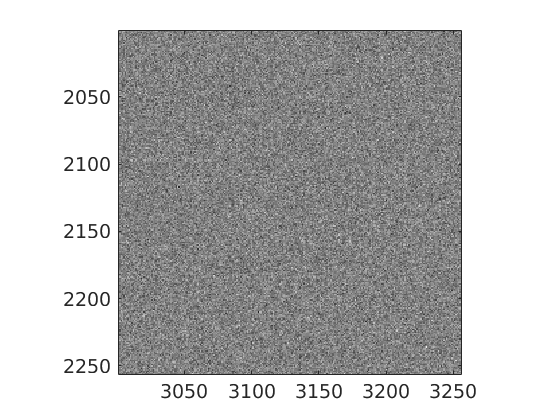}
		\caption{An excerpt of the pure noise micrograph}
	\end{subfigure} \hfill 
	\begin{subfigure}[t]{0.3\textwidth}
		\centering
		\includegraphics[scale=0.4]{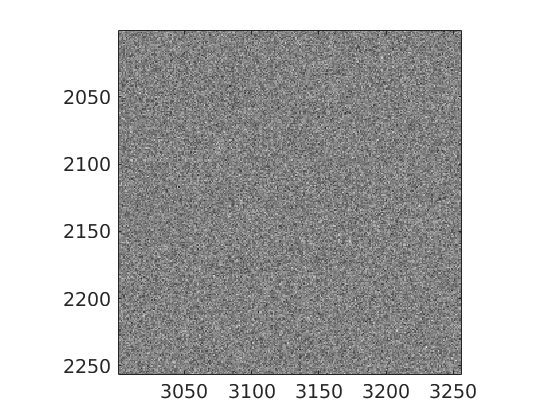}
		\caption{An excerpt of the clean micrograph in the right panel perturbed by the noise in the left panel.}
	\end{subfigure} \hfill 
	\begin{subfigure}[t]{0.3\textwidth}
		\centering
		\includegraphics[scale=0.4]{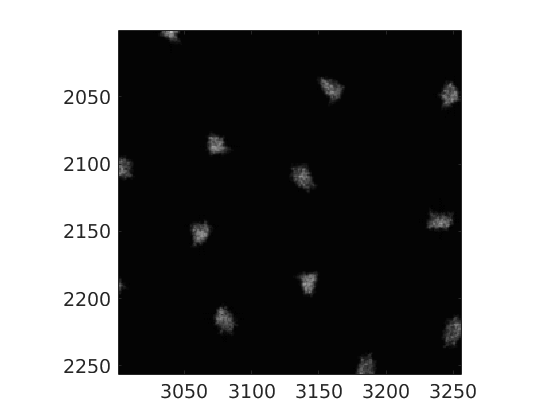}
		\caption{An excerpt of the clean micrograph}
	\end{subfigure}
	\caption{\label{fig:cryo_detection} All micrographs are of size $7420^2$ pixels and the projections are taken from the BPTI molecule of size $31^3$. The added noise was drawn from i.i.d.\ Gaussian distribution with zero mean and standard deviation $25$, corresponding to an SNR below $1/1024$.  
		The noise realization is identical in both micrographs.} 
\end{figure}

Figure~\ref{fig:cryo_detection} 
presents excerpts  of two noisy micrographs,  only one of which contains projections.
The noise level corresponds to $\text{SNR} = 1/1024$.
In the presence of projections, the estimated $\gamma$ was $0.12$, corresponding to approximately $6784$ projections. 
On the other hand, the estimated $\gamma$ drops to $10^{-5}$ for the pure noise micrograph, corresponding to less than one projection.

\paragraph{More technical details.}

	The true volume used in the experiments in Figures~\ref{fig:cryo_detection} and~\ref{fig:BPTI_recon} was the Bovine Pancreatic Trypsin Inhibitor (BPTI) mutant with altered binding loop sequence, whose atomic model is available in the Protein Data Bank (PDB)  
	as 1QLQ.\footnote{\url{https://www.rcsb.org}} We generated an EM map from this atomic model in UCSF Chimera~\cite{pettersen2004ucsf} at a resolution of 5~\AA, and cropped it to remove zeros at the boundary to obtain a volume of size $31^3$. For the experiment in Figure~\ref{fig:BPTI_recon}, the volume was downsampled to size $20^3$. For the experiment in Figure~\ref{fig:TRPV1_recon}, we used the TRPV1 in complex with DkTx and RTX, whose EM map is available in the  Electron Microscopy Data Bank (EMDB) 
	as EMD-8117.\footnote{\url{http://www.ebi.ac.uk/pdbe/emdb}} The original map has size $192^3$, and was downsampled to size $20^3$. To generate the ground truth for our reconstructions, we expanded both volumes as in~\eqref{eq:volume_expansion_app} with cutoff $L_{\text{max}} = 5$ for TRPV1 and $L_{\text{max}} = 2$ for BPTI. Each volume is described by $\sum_{\ell=0}^{L_{\max}}(2\ell+1)S(\ell)$ coefficients using the expansion~\eqref{eq:volume_expansion_app}, where the cutoffs $S(\ell)$ are determined using the Nyquist criterion as in~\cite{bhamre2017anisotropic}.

	The micrographs for the experiments  were generated as follows. 
	We sample rotation matrices from SO(3) uniformly at random using the QR-based algorithm described in~\cite{stewart1980efficient}, and generate the projection  of the volume corresponding to that rotation matrix using  ASPIRE~\cite{aspire}. The projections for the experiments were obtained from the smoothed volumes, not the original ones, to ensure that the only sources of error are the noise and our ability to invert the moments via~\eqref{eq:LS_cryo}. 
	We keep track of the indices at which the upper left corner of a projection can be placed without violating the separation condition, so all projections are separated by at least {$P-1$ pixels in each dimension, where  the projections are contained in a box of size $P\times P$.}  The location of the upper left corner of each new projection is picked uniformly at random from the set of available indices. We continue adding projections to the micrograph until no more projections can be added without violating the separation condition.

The experiments  were performed on a machine with 40 cores of Intel Xeon E5-2698 v4 @ 2.20GHz with 100 GB of RAM, and took 2 hours per reconstruction. The computation of the moments from each micrograph  was performed on a machine with 4 nVidia P100 GPUs with 16 GB of memory each and with 100 GB of RAM. It took 3 minutes per micrograph of size $7420^2$ to compute the first three autocorrelations.

\section{Discussion} \label{sec:discussion}

In this paper, we showed that, in principle, it is possible to estimate a 3-D structure from micrographs, without detecting its projections.  
Our strategy is to compute autocorrelations of the micrographs and to relate these statistics to the unknown volume's parameters. Recovering the parameters from the statistics reduces to solving a set of polynomial equations. 
Crucially, the outlined approach involves no particle picking, 
  hence  it might be possible to reconstruct small molecules, particularly, molecules that are too small to be detected in micrographs. 
In pursuing this research direction, our goal is to significantly increase the range of molecules to which cryo-EM can be successfully applied.
Concerns for model bias are also greatly reduced since no template matching is involved.
We recognize that significant challenges lay ahead for the implementation of the proposed approach to {high-resolution} 3-D reconstruction directly from the micrographs. We discuss a few now.

The numerical experiments we have performed reveal that the third-order autocorrelation may not be enough for 3-D reconstruction in practice, due to high sensitivity.
One possible remedy might be adding priors to the least squares problem~\eqref{eq:LS_cryo}. For example, one can use data-driven or sparsity aware priors~\cite{kimanius2021exploiting,shani2022denoiser,vonesch2011fast,bendory2022sparse,bendory2022autocorrelation}, or priors based on the statistical properties of typical proteins~\cite{singer2021wilson,gilles2022molecular}.   
Alternatively, this suggests that fourth-order autocorrelation may be necessary. This in turn would imply that the procedure might require a large amount of data. 
Recent trends in high-throughput cryo-EM technology   give hope that this may be a lesser concern in the long term. Still, large amounts of data also require large amounts of computation.

To reach high-resolution reconstruction, beyond data acquisition and computational challenges, there are modeling issues  to consider.
In contrast to the simplifying assumptions we have made above, the noise might be colored; the viewing directions of the particles may be distributed non-uniformly; there may be conformational heterogeneity; particles generally do not satisfy our separation condition; and micrographs undergo a contrast transfer function which we have omitted. We believe that these aspects can be handled with the same general strategy: establish a forward model relating the expected autocorrelations of the micrographs to the target volume(s) and all parameters necessary to model the above effects. 
The effect of many of these aspects was recently studied in the context of the multi-target detection  problem, which can be interpreted as a simplified model of the problem considered in this paper~\cite{bendory2019multi,marshall2020image,bendory2021multi,lan2021random}. 
For example, it was shown that the strict  separation condition can be replaced by a parameterized pair-correlation function~\cite{lan2020multi,kreymer2022two}.
Such a function models the distribution of distances between neighboring projections. The observed autocorrelations depend linearly on these parameters, which could be estimated as part of the inverse problem.
In addition, alternative computational strategies to autocorrelation analysis,  such as 
approximate expectation-maximization~\cite{lan2020multi,kreymer2021approximate}, GAN-based techniques~\cite{gupta2021cryogan}, and generalized autocorrelation analysis~\cite{abas2022generalized,weber2022generalized}, were developed. 
We believe that these techniques can be adapted to recovering molecular structures directly from the micrographs, and perhaps to alleviate some of the  computational challenges.  We hope to take care of these issues in future research. 

In addition,  our technique  allows the use of much lower defocus values (a parameter controlled by the user, and which affects the microscope's point spread function). Lower defocus means lower contrast, but also maintains  higher frequency information. Consequently, we may be able to get high resolution reconstructions from fewer micrographs.

From the information-theoretic perspective, it is essential to understand the sample complexity of the problem: how many particle projections are required to estimate the 3-D structure to a desired accuracy? In the closely-related model of multi-reference alignment, it was shown that in the finite-dimensional, low SNR regime, the sample complexity is governed by the moments~\cite{perry2019sample,bandeira2017estimation,abbe2018estimation,abbe2018multireference}. 
If the same is true for our problem, and since we believe that the third-order autocorrelation suffices to determine uniquely the 3-D structure, this will imply that $N=\omega(\sigma^6)$ is a necessary condition for accurate recovery (for a fixed $\gamma>0$). 
If the dimension of the problem is high (many parameters are required to describe the 3-D structure), the sample complexity of multi-reference alignment is no longer controlled by the moments, but by the ratio between the dimension and the noise level~\cite{romanov2021multi}; we believe a similar phenomenon will hold true for the problem studied in this paper.

\section*{Acknowledgment}
This work was primarily performed when all authors were at Princeton University. 
The authors  thank Ayelet Heimowitz, Joe Kileel, Ti-Yen Lan, Roy Lederman, Amit Moscovich, Nir Sharon, and  Fred Sigworth for helpful discussions,  Boris Landa and Yoel Shkolnisky for providing the code for the 2-D PSWFs expansion, and Alberto Bartesaghi, Amitay Eldar, and Yoel Shkolnisky  for helping to produce Figure~\ref{fig:exp_micro_example}.
The authors also thank  the reviewers for the thoughtful  and helpful comments on the manuscript.
AS was partially supported by AFOSR FA9550-20-1-0266, the Simons Foundation Math+X Investigator Award, NSF BIGDATA Award IIS-1837992, NSF DMS-2009753, and NIH/NIGMS 1R01GM136780-01.
NB was partially supported by NSF award DMS-1719558.
TB is partially supported by the BSF grant no. 2020159, 
	the NSF-BSF grant no. 2019752, and the ISF grant no. 1924/21.
WL was partially supported by BSF grant no.\ 2018230 and NSF BIGDATA Award IIS-1837992.

\bibliographystyle{plain}

\begin{thebibliography}{10}

\bibitem{aspire}
{ComputationalCryoEM/ASPIRE-Python: v0.9.2 }.
\newblock \url{https://doi.org/10.5281/zenodo.5657281}.

\bibitem{abas2022generalized}
Asaf Abas, Tamir Bendory, and Nir Sharon.
\newblock The generalized method of moments for multi-reference alignment.
\newblock {\em IEEE Transactions on Signal Processing}, 70:1377--1388, 2022.

\bibitem{abbe2018multireference}
Emmanuel Abbe, Tamir Bendory, William Leeb, Jo{\~a}o~M Pereira, Nir Sharon, and
  Amit Singer.
\newblock Multireference alignment is easier with an aperiodic translation
  distribution.
\newblock {\em IEEE Transactions on Information Theory}, 65(6):3565--3584,
  2018.

\bibitem{abbe2018estimation}
Emmanuel Abbe, Joao~M Pereira, and Amit Singer.
\newblock Estimation in the group action channel.
\newblock In {\em 2018 IEEE International Symposium on Information Theory
  (ISIT)}, pages 561--565. IEEE, 2018.

\bibitem{bai2021seeing}
Xiao-chen Bai.
\newblock Seeing atoms by single-particle cryo-{EM}.
\newblock {\em Trends in biochemical sciences}, 46(4):253--254, 2021.

\bibitem{bandeira2017estimation}
Afonso~S Bandeira, Ben Blum-Smith, Joe Kileel, Amelia Perry, Jonathan Weed, and
  Alexander~S Wein.
\newblock Estimation under group actions: recovering orbits from invariants.
\newblock {\em arXiv preprint arXiv:1712.10163}, 2017.

\bibitem{bandeira2020non}
Afonso~S Bandeira, Yutong Chen, Roy~R Lederman, and Amit Singer.
\newblock Non-unique games over compact groups and orientation estimation in
  cryo-{EM}.
\newblock {\em Inverse Problems}, 36(6):064002, 2020.

\bibitem{bartesaghi20152}
Alberto Bartesaghi, Alan Merk, Soojay Banerjee, Doreen Matthies, Xiongwu Wu,
  Jacqueline~LS Milne, and Sriram Subramaniam.
\newblock {2.2 {\AA} resolution cryo-{EM} structure of $\beta$-galactosidase in
  complex with a cell-permeant inhibitor}.
\newblock {\em Science}, 348(6239):1147--1151, 2015.

\bibitem{bendory2020single}
Tamir Bendory, Alberto Bartesaghi, and Amit Singer.
\newblock Single-particle cryo-electron microscopy: Mathematical theory,
  computational challenges, and opportunities.
\newblock {\em IEEE Signal Processing Magazine}, 37(2):58--76, 2020.

\bibitem{bendory2019multi}
Tamir Bendory, Nicolas Boumal, William Leeb, Eitan Levin, and Amit Singer.
\newblock Multi-target detection with application to cryo-electron microscopy.
\newblock {\em Inverse Problems}, 35(10):104003, 2019.

\bibitem{bendory2022autocorrelation}
Tamir Bendory, Yuehaw Khoo, Joe Kileel, Oscar Mickelin, and Amit Singer.
\newblock Autocorrelation analysis for cryo-{EM} with sparsity constraints:
  Improved sample complexity and projection-based algorithms.
\newblock {\em arXiv preprint arXiv:2209.10531}, 2022.

\bibitem{bendory2021multi}
Tamir Bendory, Ti-Yen Lan, Nicholas~F Marshall, Iris Rukshin, and Amit Singer.
\newblock Multi-target detection with rotations.
\newblock {\em arXiv preprint arXiv:2101.07709}, 2021.

\bibitem{bendory2022sparse}
Tamir Bendory, Oscar Michelin, and Amit Singer.
\newblock Sparse multi-reference alignment: Sample complexity and computational
  hardness.
\newblock In {\em ICASSP 2022-2022 IEEE International Conference on Acoustics,
  Speech and Signal Processing (ICASSP)}, pages 8977--8981. IEEE, 2022.

\bibitem{bepler2019positive}
Tristan Bepler, Andrew Morin, Micah Rapp, Julia Brasch, Lawrence Shapiro,
  Alex~J Noble, and Bonnie Berger.
\newblock Positive-unlabeled convolutional neural networks for particle picking
  in cryo-electron micrographs.
\newblock {\em Nature methods}, 16(11):1153--1160, 2019.

\bibitem{bhamre2017anisotropic}
Tejal Bhamre, Teng Zhang, and Amit Singer.
\newblock Anisotropic twicing for single particle reconstruction using
  autocorrelation analysis.
\newblock {\em arXiv preprint arXiv:1704.07969}, 2017.

\bibitem{donoho2004higher}
David Donoho and Jiashun Jin.
\newblock Higher criticism for detecting sparse heterogeneous mixtures.
\newblock {\em The Annals of Statistics}, 32(3):962--994, 2004.

\bibitem{eldar2020klt}
Amitay Eldar, Boris Landa, and Yoel Shkolnisky.
\newblock {KLT} picker: particle picking using data-driven optimal templates.
\newblock {\em Journal of structural biology}, 210(2):107473, 2020.

\bibitem{fan2021maximum}
Zhou Fan, Roy~R Lederman, Yi~Sun, Tianhao Wang, and Sheng Xu.
\newblock Maximum likelihood for high-noise group orbit estimation and
  single-particle cryo-{EM}.
\newblock {\em arXiv preprint arXiv:2107.01305}, 2021.

\bibitem{frank2006three}
Joachim Frank.
\newblock {\em Three-dimensional electron microscopy of macromolecular
  assemblies: visualization of biological molecules in their native state}.
\newblock Oxford university press, 2006.

\bibitem{frank2018single}
Joachim Frank.
\newblock Single-particle reconstruction of biological molecules—story in a
  sample ({N}obel lecture).
\newblock {\em Angewandte Chemie International Edition}, 57(34):10826--10841,
  2018.

\bibitem{gilles2022molecular}
Marc~Aur{\`e}le Gilles and Amit Singer.
\newblock A molecular prior distribution for {B}ayesian inference based on
  {W}ilson statistics.
\newblock {\em Computer Methods and Programs in Biomedicine}, page 106830,
  2022.

\bibitem{glaeser1999electron}
Robert~M Glaeser.
\newblock Electron crystallography: present excitement, a nod to the past,
  anticipating the future.
\newblock {\em Journal of structural biology}, 128(1):3--14, 1999.

\bibitem{gupta2021cryogan}
Harshit Gupta, Michael~T McCann, Laurene Donati, and Michael Unser.
\newblock {CryoGAN}: a new reconstruction paradigm for single-particle
  cryo-{EM} via deep adversarial learning.
\newblock {\em IEEE Transactions on Computational Imaging}, 7:759--774, 2021.

\bibitem{heimowitz2018apple}
Ayelet Heimowitz, Joakim And{\'e}n, and Amit Singer.
\newblock {APPLE} picker: Automatic particle picking, a low-effort cryo-{EM}
  framework.
\newblock {\em Journal of structural biology}, 204(2):215--227, 2018.

\bibitem{heimowitz2020reducing}
Ayelet Heimowitz, Joakim And{\'e}n, and Amit Singer.
\newblock Reducing bias and variance for {CTF} estimation in single particle
  cryo-{EM}.
\newblock {\em Ultramicroscopy}, 212:112950, 2020.

\bibitem{henderson1995potential}
Richard Henderson.
\newblock The potential and limitations of neutrons, electrons and {X}-rays for
  atomic resolution microscopy of unstained biological molecules.
\newblock {\em Quarterly reviews of biophysics}, 28(2):171--193, 1995.

\bibitem{henderson2013avoiding}
Richard Henderson.
\newblock Avoiding the pitfalls of single particle cryo-electron microscopy:
  {E}instein from noise.
\newblock {\em Proceedings of the National Academy of Sciences},
  110(45):18037--18041, 2013.

\bibitem{herzik2019high}
Mark~A Herzik, Mengyu Wu, and Gabriel~C Lander.
\newblock {High-resolution structure determination of sub-100 kDa complexes
  using conventional cryo-EM}.
\newblock {\em Nature communications}, 10(1):1--9, 2019.

\bibitem{iudin2016empiar}
Andrii Iudin, Paul~K Korir, Jos{\'e} Salavert-Torres, Gerard~J Kleywegt, and
  Ardan Patwardhan.
\newblock {EMPIAR}: a public archive for raw electron microscopy image data.
\newblock {\em Nature methods}, 13(5):387--388, 2016.

\bibitem{kam1980reconstruction}
Zvi Kam.
\newblock The reconstruction of structure from electron micrographs of randomly
  oriented particles.
\newblock {\em Journal of Theoretical Biology}, 82(1):15--39, 1980.

\bibitem{khoshouei2017cryo}
Maryam Khoshouei, Mazdak Radjainia, Wolfgang Baumeister, and Radostin Danev.
\newblock {Cryo-{EM} structure of haemoglobin at 3.2 {\AA} determined with the
  {V}olta phase plate}.
\newblock {\em Nature communications}, 8(1):1--6, 2017.

\bibitem{kimanius2021exploiting}
Dari Kimanius, Gustav Zickert, Takanori Nakane, Jonas Adler, Sebastian Lunz,
  C-B Sch{\"o}nlieb, Ozan {\"O}ktem, and Sjors~HW Scheres.
\newblock Exploiting prior knowledge about biological macromolecules in
  cryo-{EM} structure determination.
\newblock {\em IUCrJ}, 8(1):60--75, 2021.

\bibitem{klug1972three}
A~Klug and RA~Crowther.
\newblock Three-dimensional image reconstruction from the viewpoint of
  information theory.
\newblock {\em Nature}, 238(5365):435--440, 1972.

\bibitem{kreymer2022two}
Shay Kreymer and Tamir Bendory.
\newblock Two-dimensional multi-target detection: An autocorrelation analysis
  approach.
\newblock {\em IEEE Transactions on Signal Processing}, 70:835--849, 2022.

\bibitem{kreymer2021approximate}
Shay Kreymer, Amit Singer, and Tamir Bendory.
\newblock An approximate expectation-maximization for two-dimensional
  multi-target detection.
\newblock {\em IEEE Signal Processing Letters}, 2022.

\bibitem{lan2020multi}
Ti-Yen Lan, Tamir Bendory, Nicolas Boumal, and Amit Singer.
\newblock Multi-target detection with an arbitrary spacing distribution.
\newblock {\em IEEE Transactions on Signal Processing}, 68:1589--1601, 2020.

\bibitem{lan2021random}
Ti-Yen Lan, Nicolas Boumal, and Amit Singer.
\newblock {Random conical tilt reconstruction without particle picking in
  cryo-electron microscopy}.
\newblock {\em Acta Crystallographica Section A}, 78(4):294--301, 2022.

\bibitem{landa2017steerable}
Boris Landa and Yoel Shkolnisky.
\newblock Steerable principal components for space-frequency localized images.
\newblock {\em SIAM journal on imaging sciences}, 10(2):508--534, 2017.

\bibitem{landa2018steerable}
Boris Landa and Yoel Shkolnisky.
\newblock The steerable graph laplacian and its application to filtering image
  datasets.
\newblock {\em SIAM Journal on Imaging Sciences}, 11(4):2254--2304, 2018.

\bibitem{lander2021conquer}
Gabriel~C Lander and Robert~M Glaeser.
\newblock Conquer by cryo-{EM} without physically dividing.
\newblock {\em Biochemical Society Transactions}, 49(5):2287--2298, 2021.

\bibitem{levin20183d}
Eitan Levin, Tamir Bendory, Nicolas Boumal, Joe Kileel, and Amit Singer.
\newblock {3D} ab initio modeling in cryo-{EM} by autocorrelation analysis.
\newblock In {\em 2018 IEEE 15th International Symposium on Biomedical Imaging
  (ISBI 2018)}, pages 1569--1573. IEEE, 2018.

\bibitem{liang2017phase}
Yi-Lynn Liang, Maryam Khoshouei, Mazdak Radjainia, Yan Zhang, Alisa Glukhova,
  Jeffrey Tarrasch, David~M Thal, Sebastian~GB Furness, George Christopoulos,
  Thomas Coudrat, et~al.
\newblock Phase-plate cryo-{EM} structure of a class b {GPCR--G}-protein
  complex.
\newblock {\em Nature}, 546(7656):118--123, 2017.

\bibitem{liu2018near}
Yuxi Liu, Shane Gonen, Tamir Gonen, and Todd~O Yeates.
\newblock Near-atomic cryo-{EM} imaging of a small protein displayed on a
  designed scaffolding system.
\newblock {\em Proceedings of the National Academy of Sciences},
  115(13):3362--3367, 2018.

\bibitem{marshall2020image}
Nicholas~F Marshall, Ti-Yen Lan, Tamir Bendory, and Amit Singer.
\newblock Image recovery from rotational and translational invariants.
\newblock In {\em ICASSP 2020-2020 IEEE International Conference on Acoustics,
  Speech and Signal Processing (ICASSP)}, pages 5780--5784. IEEE, 2020.

\bibitem{natterer2001mathematics}
Frank Natterer.
\newblock {\em The mathematics of computerized tomography}.
\newblock SIAM, 2001.

\bibitem{perry2019sample}
Amelia Perry, Jonathan Weed, Afonso~S Bandeira, Philippe Rigollet, and Amit
  Singer.
\newblock The sample complexity of multireference alignment.
\newblock {\em SIAM Journal on Mathematics of Data Science}, 1(3):497--517,
  2019.

\bibitem{pettersen2004ucsf}
Eric~F Pettersen, Thomas~D Goddard, Conrad~C Huang, Gregory~S Couch, Daniel~M
  Greenblatt, Elaine~C Meng, and Thomas~E Ferrin.
\newblock {UCSF} {C}himera—a visualization system for exploratory research
  and analysis.
\newblock {\em Journal of computational chemistry}, 25(13):1605--1612, 2004.

\bibitem{romanov2021multi}
Elad Romanov, Tamir Bendory, and Or~Ordentlich.
\newblock Multi-reference alignment in high dimensions: sample complexity and
  phase transition.
\newblock {\em SIAM Journal on Mathematics of Data Science}, 3(2):494--523,
  2021.

\bibitem{scapin2018cryo}
Giovanna Scapin, Clinton~S Potter, and Bridget Carragher.
\newblock Cryo-{EM} for small molecules discovery, design, understanding, and
  application.
\newblock {\em Cell chemical biology}, 25(11):1318--1325, 2018.

\bibitem{scheres2015semi}
Sjors~HW Scheres.
\newblock Semi-automated selection of cryo-{EM} particles in {RELION}-1.3.
\newblock {\em Journal of structural biology}, 189(2):114--122, 2015.

\bibitem{schwartz2019laser}
Osip Schwartz, Jeremy~J Axelrod, Sara~L Campbell, Carter Turnbaugh, Robert~M
  Glaeser, and Holger M{\"u}ller.
\newblock Laser phase plate for transmission electron microscopy.
\newblock {\em Nature methods}, 16(10):1016--1020, 2019.

\bibitem{shani2022denoiser}
Jonathan Shani, Raja Giryes, and Tamir Bendory.
\newblock Denoiser-based projections for {2-D} super-resolution multi-reference
  alignment.
\newblock {\em arXiv preprint arXiv:2204.04754}, 2022.

\bibitem{sharon2020method}
Nir Sharon, Joe Kileel, Yuehaw Khoo, Boris Landa, and Amit Singer.
\newblock Method of moments for {3D} single particle ab initio modeling with
  non-uniform distribution of viewing angles.
\newblock {\em Inverse Problems}, 36(4):044003, 2020.

\bibitem{shatsky2009method}
Maxim Shatsky, Richard~J Hall, Steven~E Brenner, and Robert~M Glaeser.
\newblock A method for the alignment of heterogeneous macromolecules from
  electron microscopy.
\newblock {\em Journal of structural biology}, 166(1):67--78, 2009.

\bibitem{singer2021wilson}
Amit Singer.
\newblock Wilson statistics: derivation, generalization and applications to
  electron cryomicroscopy.
\newblock {\em Acta Crystallographica Section A: Foundations and Advances},
  77(5), 2021.

\bibitem{singer2020computational}
Amit Singer and Fred~J Sigworth.
\newblock Computational methods for single-particle electron cryomicroscopy.
\newblock {\em Annual Review of Biomedical Data Science}, 3:163--190, 2020.

\bibitem{slepian1964prolate}
David Slepian.
\newblock Prolate spheroidal wave functions, {F}ourier analysis and
  uncertainty{—IV}: extensions to many dimensions; generalized prolate
  spheroidal functions.
\newblock {\em Bell System Technical Journal}, 43(6):3009--3057, 1964.

\bibitem{sorzano2019survey}
Carlos Oscar~S Sorzano, A~Jim{\'e}nez, Javier Mota, Jos{\'e}~Luis Vilas, David
  Maluenda, Marta Mart{\'\i}nez, Erney Ram{\'\i}rez-Aportela, Tomas Majtner,
  J~Segura, Ruben S{\'a}nchez-Garc{\'\i}a, et~al.
\newblock Survey of the analysis of continuous conformational variability of
  biological macromolecules by electron microscopy.
\newblock {\em Acta Crystallographica Section F: Structural Biology
  Communications}, 75(1):19--32, 2019.

\bibitem{stewart1980efficient}
Gilbert~W Stewart.
\newblock The efficient generation of random orthogonal matrices with an
  application to condition estimators.
\newblock {\em SIAM Journal on Numerical Analysis}, 17(3):403--409, 1980.

\bibitem{van2013finding}
Marin van Heel.
\newblock Finding trimeric {HIV-1} envelope glycoproteins in random noise.
\newblock {\em Proceedings of the National Academy of Sciences},
  110(45):E4175--E4177, 2013.

\bibitem{van1992correlation}
Marin van Heel, Michael Schatz, and Elena Orlova.
\newblock Correlation functions revisited.
\newblock {\em Ultramicroscopy}, 46(1-4):307--316, 1992.

\bibitem{vonesch2011fast}
C{\'e}dric Vonesch, Lanhui Wang, Yoel Shkolnisky, and Amit Singer.
\newblock Fast wavelet-based single-particle reconstruction in cryo-{EM}.
\newblock In {\em 2011 IEEE International Symposium on Biomedical Imaging: From
  Nano to Macro}, pages 1950--1953. IEEE, 2011.

\bibitem{weber2022generalized}
Ran Weber, Asaf Abas, Shay Kreymer, Tamir Bendory, et~al.
\newblock Generalized autocorrelation analysis for multi-target detection.
\newblock In {\em ICASSP 2022-2022 IEEE International Conference on Acoustics,
  Speech and Signal Processing (ICASSP)}, pages 5907--5911. IEEE, 2022.

\bibitem{wong2014cryo}
Wilson Wong, Xiao-chen Bai, Alan Brown, Israel~S Fernandez, Eric Hanssen,
  Melanie Condron, Yan~Hong Tan, Jake Baum, and Sjors~HW Scheres.
\newblock {Cryo-EM structure of the Plasmodium falciparum 80S ribosome bound to
  the anti-protozoan drug emetine}.
\newblock {\em Elife}, 3:e03080, 2014.

\bibitem{wu2020low}
Mengyu Wu and Gabriel~C Lander.
\newblock How low can we go? structure determination of small biological
  complexes using single-particle cryo-{EM}.
\newblock {\em Current opinion in structural biology}, 64:9--16, 2020.

\bibitem{zhang2019cryo}
Kaiming Zhang, Shanshan Li, Kalli Kappel, Grigore Pintilie, Zhaoming Su,
  Tung-Chung Mou, Michael~F Schmid, Rhiju Das, and Wah Chiu.
\newblock Cryo-{EM} structure of a 40 {kDa} {SAM-IV} riboswitch {RNA} at 3.7
  {\aa} resolution.
\newblock {\em Nature communications}, 10(1):1--6, 2019.

\bibitem{zhao2016fast}
Zhizhen Zhao, Yoel Shkolnisky, and Amit Singer.
\newblock Fast steerable principal component analysis.
\newblock {\em IEEE Transactions on Computational Imaging}, 2(1):1--12, 2016.

\end{thebibliography}

\appendix

\section{Proof of Proposition~\ref{prop:two_gauss}} \label{sec:proof_two_gauss}

The proof is based on  a variant of the Neyman-Pearson Lemma to derive the best (deterministic) estimator $\hat{\eta}$. Take any deterministic estimator $\hat{\eta} = \hat{\eta}(X)$, with values in $\{0,1\}$; then $\hat{\eta}$ is characterized by $S$, defined to be the set of $X$'s where $\hat{\eta} = 1$.
We write $\P_i$ to mean the probability conditional on the event $\eta = i$; that is, $\P_i[A] = \P[A | \eta = i]$. Then, the probability that $\hat{\eta}$ fails is:
\begin{align}
\label{eq:neyman-pearson}
\P[\hat{\eta} \ne \eta] 
&= q \P_0[\hat{\eta} = 1 ] + (1-q) \P_1[\hat{\eta} = 0]
\nonumber \\
&= q \P_0[\hat{\eta} = 1] + (1-q) (1 - \P_1[\hat{\eta} = 1])
\nonumber \\
&= q \P_0[\hat{\eta} = 1] + (1-q) - (1-q) \P_1[\hat{\eta} = 1]
\nonumber \\
&= (1-q)  + \int_S (q f_0(x) - (1-q) f_1(x))dx,
\end{align}
where $f_i(x)$ is the normal density with mean $\theta_i$ and variance $\sigma^2$. 
The best estimator of $\eta$ based on $X$ minimizes the failure probability; hence, it minimizes the integral in~\eqref{eq:neyman-pearson} through an appropriate choice of the set $S$. This is achieved by picking all $x$'s such that the integrand is nonpositive:
\begin{align*}
S & = \left\{x : qf_0(x) - (1-q)f_1(x) \leq 0 \right\}.
\end{align*}
With $\Lambda(x) = f_0(x) / f_1(x)$ and $b = (1-q) / q$, the corresponding estimator is:
\begin{align*}
\hat{\eta} = 
\begin{cases}
1  \text{ if } \Lambda(x) \leq b, \\
0  \text{ if } \Lambda(x) > b.
\end{cases}
\end{align*}
Taking logarithms, the set $S$ can be rewritten as the set of $x$'s where:
\begin{align*}
-\| x - \theta_0\|^2 \le - \|x - \theta_1\|^2 + 2 \sigma^2 \log(b),
\end{align*}
or equivalently
\begin{align*}
\langle x , \theta_1 - \theta_0 \rangle 
\ge \frac{\|\theta_1\|^2 - \|\theta_0\|^2}{2} - \sigma^2 \log(b).
\end{align*}
Now let us compute the probability of failure conditional on the event $\eta = 0$. In this case, failure occurs when $X \in S$. Since $X|(\eta=0) \sim N(\theta_0,\sigma^2)$, we can write $X | (\eta=0) = \sigma Z + \theta_0$, where $Z \sim N(0,I)$. On that condition,
\begin{align*}
\langle X , \theta_1 - \theta_0 \rangle 
&= \sigma \langle Z , \theta_1 - \theta_0 \rangle 
+ \langle \theta_0,\theta_1 - \theta_0 \rangle
\nonumber \\
&= \sigma \langle Z , \theta_1 - \theta_0 \rangle 
+ \langle \theta_0,\theta_1 \rangle - \|\theta_0\|^2,
\end{align*}
and failure occurs when 
\begin{align*}
\sigma\langle Z, \theta_1 - \theta_0 \rangle 
\ge \frac{\|\theta_1\|^2 + \|\theta_0\|^2}{2} - \langle \theta_0,\theta_1 \rangle
- \sigma^2 \log(b)
= \frac{1}{2} \|\theta_1 - \theta_0\|^2 - \sigma^2 \log(b).
\end{align*}
Define $Y = \langle Z , \theta_1 - \theta_0 \rangle \sim N(0,\|\theta_1 - \theta_0\|^2)$ and divide through by $\sigma$. The above event is equivalent to:
\begin{align*}
Y \ge \frac{c}{\sigma} - \sigma \log(b),
\end{align*}
where $c = \|\theta_1 - \theta_0\|^2 / 2$. 
For simplicity, let us assume $\| \theta_1 - \theta_0\| = 1$, so that $Y \sim N(0,1)$. Then, 
\begin{align*}
\P_0[\hat{\eta} = 1] = \P\left[Y \ge \frac{c}{\sigma} - \sigma \log(b) \right],
\quad Y \sim N(0,1).
\end{align*}
Similarly,
\begin{align*}
\P_1[\hat{\eta} = 0] = \P\left[Y \ge \frac{c}{\sigma} + \sigma \log(b) \right],
\quad Y \sim N(0,1).
\end{align*}
Thus, the overall probability of failure is:
%
\begin{equation*}
\P[\hat{\eta} \ne \eta] = q \P\left[Y \ge \frac{c}{\sigma} - \sigma \log(b) \right] 
+ (1-q) \P\left[Y \ge \frac{c}{\sigma} + \sigma \log(b) \right].
\end{equation*}
Now, if $q = 1/2$, then $\log(b) = 0$. Hence the probability of failure is simply:
\begin{align*}
\P\left[Y \ge \frac{c}{\sigma} \right] \longrightarrow \frac{1}{2} = q
\quad \text{as} \quad \sigma \to \infty.
\end{align*}
If $q > 1/2$, then $q > 1-q$ and $\log(b) < 0$. Consequently, 
\begin{align*}
\P\left[Y \ge \frac{c}{\sigma} - \sigma \log(b) \right] \longrightarrow 0,
\end{align*}
while
\begin{align*}
\P\left[Y \ge \frac{c}{\sigma} + \sigma \log(b) \right] \longrightarrow 1,
\end{align*}
as $\sigma \to \infty$. Hence,
\begin{align*}
\P[\hat{\eta} \ne \eta]
& \longrightarrow 1-q  \quad \text{as} \quad  \sigma \to \infty.
\end{align*}
That is, the probability of success converges to $q$.
Finally, if $q < 1/2$, then $\log(b) > 0$ and a similar reasoning shows the probability of success converges to $1-q$.
In all cases, the probability of success of the best possible deterministic estimator converges to $\max(q, 1-q)$.

\end{document}